\documentclass[10pt]{article}
\usepackage{times,amsmath,amsthm,amssymb,graphicx,xspace,epsfig,xcolor}
\usepackage{algorithm}\usepackage{algorithmic}
\usepackage{tikz, tkz-graph, tkz-berge}
\usepackage{pgf}
\usepackage{pgffor}
\usepackage{xspace}
\usepackage{color}
\usepackage{comment}
\usepackage{authblk}
\usepackage{multirow}
\usepackage{hyperref}
\usepackage{cleveref}
\usepackage{dsfont}
\usepackage{enumerate}

\usepackage{thm-restate}

\usepackage{caption}
\usepackage{subcaption}
\usepackage{soul}

\newtheorem{theorem}{Theorem}
\newtheorem{corollary}[theorem]{Corollary}
\newtheorem{proposition}[theorem]{Proposition}
\newtheorem{lemma}[theorem]{Lemma}

\theoremstyle{definition}
\newtheorem{remark}[theorem]{Remark}

\newtheorem{problem}[theorem]{Problem}
\newtheorem{conjecture}[theorem]{Conjecture}

\newcommand{\defproblem}[3]{
 \vspace{1mm}
\noindent\fbox{
 \begin{minipage}{0.96\textwidth}
 \begin{tabular*}{\textwidth}{@{\extracolsep{\fill}}lr} #1  \\ \end{tabular*}
 {\bf{Input:}} #2 \\
 {\bf{Question:}} #3
 \end{minipage}
 }
 \vspace{1mm}
}

\newcounter{claim}
\counterwithin{claim}{theorem}

\newenvironment{claim}[1][]
{\refstepcounter{claim}\vspace{1ex}\noindent{\bf Claim~\theclaim : }\it}{\vspace{1ex}}

\newenvironment{proofclaim}[1][]
	{\par\noindent {\it Proof of claim}. }{ \hfill$\lozenge$\par\vspace{11pt}}

\newenvironment{subproof}{\par\noindent {\it Proof of claim}.\ }{\hfill$\lozenge$\par\vspace{11pt}}

\DeclareMathOperator{\Mad}{Mad}

\DeclareMathOperator{\diff}{diff}

\newcommand{\dic}{\vec{\chi}}
\newcommand{\bid}{\overleftrightarrow}

\begin{document}

\title{Digraph redicolouring \thanks{Research supported by research grant
    DIGRAPHS ANR-19-CE48-0013 and by the French government, through the EUR DS4H Investments in the Future project managed by the National Research Agency (ANR) with the reference number ANR-17-EURE-0004.}}

\author{N. Bousquet$^1$, F. Havet$^2$, N. Nisse$^2$, L. Picasarri-Arrieta$^2$, A. Reinald$^{2,3}$}
\date{}

\maketitle
\vspace{-10mm}
\begin{center}
{\small 
$^1$ LIRIS, CNRS, Université Claude Bernard Lyon 1, Lyon, France\\
\texttt{nicolas.bousquet@univ-lyon1.fr}\\[3mm] 
$^2$ CNRS, Université Côte d'Azur, I3S, Inria, Sophia-Antipolis, France\\
\texttt{$\{$frederic.havet,nicolas.nisse,lucas.picasarri-arrieta$\}$@inria.fr}\\[3mm]
$^3$ LIRMM, CNRS, Université de Montpellier, Montpellier, France\\
\texttt{amadeus.reinald@lirmm.fr}\\
}
\end{center}

\affil[1]{CNRS, Universit\'e C\^ote d'Azur, I3S, INRIA, Sophia Antipolis, France}
\affil[2]{CNRS, LIRIS}
\affil[3]{LIRMM, Université de Montpellier, CNRS, Montpellier, France}

\maketitle

\begin{abstract}
    
    In this work, we generalize several results on graph recolouring to digraphs. 
    
    Given two $k$-dicolourings of a digraph $D$, we prove that it is PSPACE-complete to decide whether we can transform one into the other by recolouring one vertex at each step while maintaining a dicolouring at any step even for $k=2$ and for digraphs with maximum degree $5$ or oriented planar graphs with maximum degree $6$.

    A digraph is said to be $k$-mixing if there exists a transformation between any pair of $k$-dicolourings.
    We show that every digraph $D$ is $k$-mixing for all $k\geq \delta^*_{\min}(D)+2$, generalizing a result due to Dyer et al. We also prove that every oriented graph $\vec{G}$ is $k$-mixing for all $k\geq \delta^*_{\max}(\vec{G}) +1$ and for all $k\geq \delta^*_{\rm avg}(\vec{G})+1$. Here $\delta^*_{\min}$, $\delta^*_{\max}$, and $\delta^*_{\rm avg}$ denote the min-degeneracy, the max-degeneracy, and the average-degeneracy respectively.
    
    We pose as a conjecture that, for every digraph $D$, the dicolouring graph of $D$ on $k\geq \delta_{\min}^*(D)+2$ colours has diameter at most $O(|V(D)|^2)$. This is the analogue of Cereceda's conjecture for digraphs. We generalize to digraphs two results supporting Cereceda's conjecture. We first prove that the dicolouring graph of any digraph $D$ on $k\geq 2\delta_{\min}^*(D) + 2$ colours has linear diameter, extending a result from Bousquet and Perarnau. We also prove that the analogue of Cereceda's conjecture is true when $k\geq \frac{3}{2}(\delta_{\min}^*(D)+1)$, which generalizes a result from Bousquet and Heinrich.
    
    Restricted to the special case of oriented graphs, we prove that the dicolouring graph of any subcubic oriented graph on $k\geq 2$ colours is connected and has diameter at most $2n$. We conjecture that every non $2$-mixing oriented graph has maximum average degree at least $4$, and we provide some support for this conjecture by proving it on the special case of $2$-freezable oriented graphs. More generally, we show that every $k$-freezable oriented graph on $n$ vertices must contain at least $kn + k(k-2)$ arcs, and we give a family of $k$-freezable oriented graphs that reach this bound. In the general case, we prove as a partial result that every non $2$-mixing oriented graph has maximum average degree at least $\frac{7}{2}$.
 
\noindent{} {\bf Keywords:} Digraphs, oriented graphs, graphs recolouring, reconfiguration, dicolouring
\end{abstract}

\bibliographystyle{plain}

\begin{sloppypar}

\section{Introduction}
\label{section:introduction}

\subsection{Graph recolouring}

We denote by $[k]$ the set $\{1, \dots , k\}$. 
 Given a graph $G=(V,E)$, a \textit{$k$-colouring} of $G$ is a function $\alpha: V \xrightarrow{} [k]$ such that $\alpha(x) \neq \alpha(y)$ for every edge $xy\in E$.
 The \textit{chromatic number} $\chi(G)$ is the smallest $k$ such that $G$ admits a $k$-colouring.
 For any $k\geq \chi(G)$, the \textit{$k$-colouring graph} of $G$, denoted by ${\cal C}_k(G)$, is the graph whose vertices are the $k$-colourings of $G$ and in which two $k$-colourings are adjacent if they differ on exactly one vertex.
 We identify a path between two given colourings in ${\cal C}_k(G)$ with a {\it sequence of recolourings}, that is, an ordered list of pairs composed of a vertex of $G$ and a new colour for this vertex.
 If ${\cal C}_k(G)$ is connected, we say that $G$ is \textit{$k$-mixing}. 
 Given a graph $G$, one may ask for which values of $k$ it is $k$-mixing, and when it is, how many steps are required at most to get from one colouring to another, i.e. what is the diameter of ${\cal C}_k(G)$?

 Determining if a graph is $k$-mixing has applications in statistical physics, where colourings represent states of the antiferromagnetic Potts model at temperature zero. The questions above were first addressed by researchers studying the Glauber dynamics for sampling $k$-colourings of a given graph. This is a Markov chain used to obtain efficient algorithms for approximately counting or almost uniformly sampling $k$-colourings of a graph, and the connectedness of the $k$-colouring graph is a necessary condition for such a Markov chain to be rapidly mixing. 
 In graph theory, the study of recolouring has been rapidly developing in the last fifteen years, since the works of Cereceda, van den Heuvel and Johnson~\cite{cerecedaEJC30,cerecedaJGT67}.
 We refer the reader to the PhD thesis of Bartier~\cite{bartierTHESIS} for a complete overview on graph recolouring and to the surveys of van Heuvel~\cite{heuvel13} and Nishimura~\cite{Nishimura18} for reconfiguration problems in general. 
 
As described above, one of the main problems in recolouring is to decide whether a given graph is $k$-mixing. \\
\defproblem{\textsc{Is $k$-Mixing}}{A graph $G$}{Is $G$ $k$-mixing?}

Note that {\sc Is $2$-Mixing} is trivial since only edgeless graphs are $2$-mixing.
On the other hand, Cereceda~\cite{cerecedaEJC30} proved that {\sc Is $3$-Mixing} is coNP-complete. For higher values of $k$, the complexity of this problem is still open.
 
A related problem is that of recognizing when two given $k$-colourings of a graph $G$ are in the same connected component of ${\cal C}_k(G)$. Formally, we have the following decision problem: \\
\defproblem{\textsc{$k$-Colouring-Path}}{A graph $G$ along with two $k$-colourings $\alpha$ and $\beta$ of $G$.}{Is there a path between $\alpha$ and $\beta$ in ${\cal C}_k(G)$?}

{\sc $2$-Colouring Path} is trivial since only isolated vertices can be recoloured in a bipartite graph.
Cereceda, van den Heuvel and Johnson~\cite{cerecedaJGT67} proved that {\sc $3$-Colouring Path} is polynomial-time solvable.
Moreover the authors proved that the diameter of each component of ${\cal C}_k(G)$ is $O(n^2)$.
 
In contrast, for every $k\geq 4$, Bonsma and Cereceda~\cite{bonsmaTCS410} showed a family ${\cal G}_k$ of graphs such that for every $G\in {\cal G}_k$ of order $n$, there exists two $k$-colourings whose distance in ${\cal C}_k(G)$ is finite and superpolynomial in $n$.
They also proved that {\sc $k$-Colouring Path} is PSPACE-complete for all $k\geq 4$ even restricted to bipartite graphs.
However, the situation is different for degenerate graphs.
The \textit{degeneracy} of a graph $G$, denoted by $\delta^*(G)$, is the largest minimum degree of any subgraph of $G$. 
Bonsma and Cereceda~\cite{bonsmaTCS410}  and  Dyer et al.~\cite{dyerRSA29} independently proved the following.
\begin{theorem}[Bonsma and Cereceda~\cite{bonsmaTCS410} ; Dyer et al.~\cite{dyerRSA29}]
    \label{thm:bonsma-cereda-dyver}
    Let $k\in \mathbb{N}$ and $G$ be a graph.
    
    If $k\geq \delta^*(G) +2$, then $G$ is $k$-mixing. 
\end{theorem}
Cereceda~\cite{cerecedaTHESIS} also conjectured the following.
\begin{conjecture}[Cereceda~\cite{cerecedaTHESIS}]
    \label{conj:cereceda}
    Let $k\in \mathbb{N}$ and $G$ be a graph.
    
    If $k\geq \delta^*(G)+2$, then the diameter of ${\cal C}_k(G)$ is at most $O(n^2)$.
\end{conjecture}
Cereceda~\cite{cerecedaTHESIS} proved that this is true when $k \geq 2\delta^*(G)+1$. This was improved recently by Bousquet and Heinrich~\cite{bousquetHeinrichCoRR}, who showed the following.
\begin{theorem}[Bousquet and Heinrich~\cite{bousquetHeinrichCoRR}]
    \label{thm:bousquet-heinrich}
    Let $k\in \mathbb{N}$ and $G$ be a graph. Then ${\cal C}_k(G)$ has diameter at most:
    \begin{itemize}
        \item $Cn^2$ if $k\geq \frac{3}{2}(\delta^*(G) +1)$ (where $C$ is a constant independent from $k$),
        \item $C_{\varepsilon}n^{\lceil \frac{1}{\varepsilon} \rceil}$ if $k\geq (1+\varepsilon)(\delta^*(G) +2)$ (where $C_\varepsilon$ is a constant independent from $k$),
        \item $(Cn)^{\delta^*(G)+1}$ for any $k\geq \delta^*(G) +2$ (where $C$ is a constant independent from $k$).
    \end{itemize}
\end{theorem}
Bousquet and Perarnau~\cite{bousquetEJC52} also proved the following.
\begin{theorem}[Bousquet and Perarnau~\cite{bousquetEJC52}]
\label{thm:bousquet-perarnau}
    Let $k\in \mathbb{N}$ and $G$ be a graph.
    
    If $k\geq 2\delta^*(G)+2$, then the diameter of ${\cal C}_k(G)$ is at most $(\delta^*(G)+1)n$.
\end{theorem}

\medskip

Let $k\in \mathbb{N}$ and $G$ be a graph that is not $k$-mixing. It follows from Theorem~\ref{thm:bonsma-cereda-dyver} that $G$ contains a subgraph $H$ with minimum degree at least $k-1$. Thus $G$ has maximum average degree at least $k-1$. This bound is tight because the complete graph on $k$ vertices is $(k-1)$-regular and is not $k$-mixing. Moreover, it is shown in~\cite{bonamyENDM68} that this bound is tight even when we restrict to graphs of arbitrary large girth. The initial proof uses the probabilistic method. In Section~\ref{section:density}, we give a new constructive proof of this result, based on an explicit construction of regular bipartite graphs from Lazebnik and Ustimenko~\cite{lazebnikDAM60}.

\subsection{Digraph redicolouring.}

Let $D$ be a digraph. A \textit{digon} is a pair of arcs in opposite directions between the same vertices. 
An \textit{oriented graph} is a digraph with no digon. The \textit{bidirected graph} associated to a graph $G$, denoted by $\bid{G}$,  is the digraph obtained from $G$, by replacing every edge by a digon. A \textit{tournament} on $k$ vertices is an orientation of the complete graph on $k$ vertices. The \textit{transitive tournament} on $k$ vertices is the only acyclic tournament on $k$ vertices.

In 1982, Neumann-Lara~\cite{neumannlaraJCT33} introduced the notions of dicolouring and dichromatic number, which generalize the ones of colouring and chromatic number.
A \textit{$k$-dicolouring} of $D$ is a function $\alpha: V(D) \rightarrow{} [k]$ such that $\alpha^{-1}(i)$ induces an acyclic subdigraph in $D$ for each $i \in [k]$. The \textit{dichromatic number} of $D$, denoted by $\dic(D)$, is the smallest $k$ such that $D$ admits a $k$-dicolouring. 
There is a one-to-one correspondence between the $k$-colourings of a graph $G$ and the $k$-dicolourings of the associated bidirected graph $\bid{G}$, and in particular $\chi(G) = \dic(\bid{G})$.
Hence every result on graph colourings can be seen as a result on dicolourings of bidirected graphs, and it is natural to study whether the result can be extended to all digraphs.

In this paper, we study digraph redicolouring, which is a generalization of graph recolouring.
For any $k\geq \dic(D)$, the \textit{$k$-dicolouring graph} of $D$, denoted by ${\cal D}_k(D)$, is the graph whose vertices are the $k$-dicolourings of $D$ and in which two $k$-dicolourings are adjacent if they differ on exactly one vertex.  Observe that ${\cal C}_k(G) = {\cal D}_k(\bid{G})$ for any bidirected graph $\bid{G}$.  A \textit{redicolouring sequence} between two dicolourings is a path between these dicolourings in ${\cal D}_k(D)$.  The digraph $D$ is \textit{$k$-mixing} if ${\cal D}_k(D)$ is connected, and \textit{$k$-freezable} if  ${\cal D}_k(D)$ contains an isolated vertex. A vertex $v$ is \textit{blocked} to its colour in a $k$-dicolouring $\alpha$ if, for every colour $c \in [k]$ different from $\alpha(v)$, recolouring $v$ to $c$ in $\alpha$ creates a monochromatic cycle. We say that $v$ is \emph{frozen} in $\alpha$ if $\beta(v) = \alpha(v)$ for any $k$-dicolouring $\beta$ in the same connected component of $\alpha$ in ${\cal D}_k(D)$.

    \medskip
In Section~\ref{section:complexity}, we consider the directed analogues of {\sc Is $k$-Mixing} and {\sc $k$-Colouring Path}.

\medskip

\defproblem{\sc Directed Is $k$-Mixing}{A digraph $D$.}{Is $D$ $k$-mixing?}

\defproblem{\sc $k$-Dicolouring Path}{A digraph $D$ along with two $k$-dicolourings $\alpha$ and $\beta$ of $D$.}{Is there a path between $\alpha$ and $\beta$ in ${\cal D}_k(D)$?}

Note that {\sc Is $k$-Mixing} and {\sc $k$-Dicolouring Path} may be seen as the restrictions of {\sc Directed Is $k$-Mixing} and {\sc $k$-Dicolouring Path} to bidirected graphs.
Therefore hardness results transfer to those problems.
It follows that {\sc Directed Is $3$-Mixing} is NP-hard and {\sc $k$-Dicolouring Path} is PSPACE-complete for all $k\geq 4$.
We strengthen these results is Section~\ref{section:complexity} by proving that {\sc $2$-Dicolouring Path} is PSPACE-complete, and that 
{\sc $k$-Dicolouring Path} remains PSPACE-complete when restricted to some digraph classes:
\begin{restatable}{theorem}{pspacecompleteness}\label{thm:pspace-completeness}
\begin{enumerate}[(i)]
    \item For every $k \ge 2$, {\sc $k$-Dicolouring Path} is PSPACE-complete on digraphs with maximum degree $2k+1$.
    \item For every $k\geq 2$, {\sc $k$-Dicolouring Path} is PSPACE-complete on oriented graphs.
    \item For every $2 \le k \le 4$, {\sc $k$-Dicolouring Path} is PSPACE-complete on planar digraphs with maximum degree $2k+2$.
    \item {\sc $2$-Dicolouring Path} is PSPACE-complete on planar oriented graphs of degree at most $6$.
\end{enumerate}
\end{restatable}

\medskip 

In Section~\ref{section:diameter}, we consider generalizations of Theorems~\ref{thm:bonsma-cereda-dyver},~\ref{thm:bousquet-heinrich} and~\ref{thm:bousquet-perarnau} to digraphs.
There are several notions of degeneracy for digraphs.
The \textit{min-degeneracy} of $D$ is  $ \delta^*_{\min}(D) = \max\{\min\{\delta^+(H), \delta^-(H)\}  \mid \mbox{$H$ subdigraph of $D$}\}$.
It is the smallest $k$ such that every subdigraph $H$ of $D$ has a vertex $v$ with $\min \{d^+_H(v), d^-_H(v)\}\leq k$.
The \textit{out-degeneracy} of $D$ is  $ \delta^*_{\rm out}(D) = \max\{\delta^+(H)  \mid \mbox{$H$ subdigraph of $D$}\}$. It is the smallest $k$ such that every subdigraph $H$ of $D$ has $\delta^+(H)\leq k$.
The \textit{max-degeneracy} of $D$, denoted by $ \delta^*_{\max}(D)$, is the smallest $k$ such that every subdigraph $H$ of $D$ has a vertex $v$ with $\max \{d^+_H(v), d^-_H(v)\}\leq k$.
The \textit{average-degeneracy} of $D$, denoted by $\delta^*_{\rm avg}(D)$, is the smallest $k$ such that every subdigraph $H$ of $D$ has a vertex $v$ with $\frac{1}{2}(d^+_H(v)+d^-_H(v))\leq k$. (In that case $k$ may be half-integral.)
Note that if $D$ is an oriented graph, then its average-degeneracy is half the degeneracy of its underlying graph.

By definition, we have 
$ \delta^*_{\min}(D) \leq \delta^*_{\rm out}(D) \leq \delta^*_{\max}(D)$
and $ \delta^*_{\min}(D) \leq \delta^*_{\rm avg}(D) \leq \delta^*_{\max}(D)$.
If $\bid{G}$ is a bidirected graph, then all directed versions of degeneracy above are equal to the degeneracy of the associated graph: 
$$\delta^*(G) = \delta^*_{\min}(\bid{G}) =  \delta^*_{\rm out}(\bid{G}) = \delta^*_{\max}(\bid{G}) = \delta^*_{\rm avg}(\bid{G}).$$

We show the following theorem, which extends Theorem~\ref{thm:bonsma-cereda-dyver} for min-degeneracy and thus also for out- , max- and average-degeneracy by the above inequalities.
\begin{restatable}{theorem}{mindegen}
\label{thm:min-degen}
 Every digraph $D$ is $k$-mixing for every $k\geq \delta^*_{\min}(D)+2$.
\end{restatable}

In the special case of oriented graphs, we improve the lower bounds on $k$ by $1$ in terms of the max-degeneracy and average-degeneracy.

\begin{restatable}{theorem}{avgdegen}\label{thm:avg-degen}
 Every oriented graph $\vec{G}$ is $k$-mixing for all $k\geq \left\lceil \delta^*_{\rm avg}(\vec{G}) \right\rceil +1$.
\end{restatable}

We also show that such an improvement cannot hold for out-degeneracy (and thus min-degeneracy) by exhibiting $k$-out-degenerate oriented graphs that are not $(k+1)$-mixing (Proposition~\ref{prop:no-improv}).

\medskip

A natural question is then the generalization of Conjecture~\ref{conj:cereceda} to digraphs.
\begin{conjecture}
Let $D$ be a digraph on $n$ vertices. 
If $k\geq \delta^*_{\min}(D) + 2$, then the diameter of ${\cal D}_k(D)$ is at most $O(n^2)$.
\end{conjecture}

As evidence towards this conjecture, we establish several results.
We first prove the analogue of Theorem~\ref{thm:bousquet-perarnau} for digraphs.
\begin{restatable}{theorem}{lineardiam}
\label{thm:linear-diameter}
Let $D$ be a digraph on $n$ vertices and $k\geq 2\delta^*_{\min}(D)+2$ be an integer.
Then, the diameter of ${\cal D}_k(D)$ is at most $(\delta^*_{\min}(D)+1)n$.
\end{restatable}

 We also generalize the first point of Theorem~\ref{thm:bousquet-heinrich}.

 \begin{restatable}{theorem}{quaddiam}
\label{thm:quadratic-diameter}
For a digraph $D$ on $n$ vertices, and $k\geq \frac{3}{2}(\delta^*_{\min}(D)+1)$, the diameter of ${\cal D}_k(D)$ is at most $Cn^2$ where $C$ is independent from $k$.
\end{restatable}

We also consider oriented subcubic graphs.
A \textit{subcubic graph} is a graph with maximum degree at most $3$. We prove that any orientation of such a graph is $2$-mixing, and that its $2$-dicolouring graph has linear diameter.

\begin{restatable}{theorem}{subcubic}\label{thm:subcubic}
Let $\vec{G}$ be a subcubic oriented graph of order $n$. Then ${\cal D}_2(\vec{G})$ is connected and has diameter at most $2n$.
\end{restatable}

\begin{figure}[hbtp]
        \begin{minipage}{\linewidth}
            \begin{center}	
              \begin{tikzpicture}[thick,scale=1, every node/.style={transform shape}]
        	    \tikzset{vertex/.style = {circle,fill=black,minimum size=5pt,
                                    inner sep=0pt}}
                \tikzset{edge/.style = {->,> = latex'}}
                \node[vertex] (a) at  (-2,0) {};
                \node[vertex,red] (b) at  (0,2) {};
                \node[vertex] (c) at  (2,0) {};                
                \node[vertex,red] (d) at  (0,-2) {};        
                \node[vertex] (e) at  (-1,0) {};                
                \node[vertex,red] (f) at  (0,1) {};
                \node[vertex] (g) at  (1,0) {};                
                \node[vertex,red] (h) at  (0,-1) {};
                
                \draw[edge] (a) to (b);
                \draw[edge] (b) to (c);
                \draw[edge] (c) to (d);
                \draw[edge] (d) to (a);
                
                \draw[edge] (e) to (f);
                \draw[edge] (f) to (g);
                \draw[edge] (g) to (h);
                \draw[edge] (h) to (e);
                
                \draw[edge] (f) to (a);
                \draw[edge] (g) to (b);
                \draw[edge] (h) to (c);
                \draw[edge] (e) to (d);
                
                \draw[edge] (a) to (e);
                \draw[edge] (b) to (f);
                \draw[edge] (c) to (g);
                \draw[edge] (d) to (h);
            \end{tikzpicture}
            \caption{An orientation of a 4-regular oriented graph with a frozen $2$-colouring.}
        \label{fig:2diregular_2frozen}
        \end{center}    
    \end{minipage}
\end{figure}
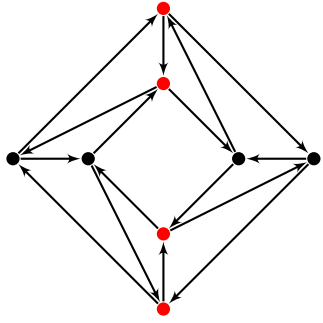

This result is tight because there exists orientations of $4$-regular graphs which are $2$-freezable (see Figure~\ref{fig:2diregular_2frozen}).

Since any digraph containing a digon is not $2$-mixing,  Theorem~\ref{thm:subcubic} directly implies the following:

\begin{corollary}\label{cor:subcubic}
A subcubic digraph $D$ is $2$-mixing if and only if $D$ is an oriented graph.
\end{corollary}

\medskip

In Section~\ref{section:density}, we turn our focus to the density of non-mixing graphs and digraphs. In a first part, we first consider undirected graphs. 
The \textit{maximum average degree} of a graph $G$, denoted by  $\Mad(G)$, is defined by $\Mad(G)=\max\left\{  \frac{2|E(H)|}{|V(H)|}\mid \mbox{$H$ a non-empty subgraph of $G$}\right\}$. In the undirected case, one can easily deduce from Theorem~\ref{thm:bonsma-cereda-dyver} that any non $k$-mixing graph $G$ contains a subgraph $H$ with minimum degree at least $k-1$. This bound is tight because the complete graph on $k$ vertices is $(k-1)$-regular and is not $k$-mixing. Using probabilistic arguments, Bonamy, Bousquet and Perarnau~\cite{bonamyENDM68} showed that this bound is tight even on graphs of arbitrary large girth (Recall that the  \textit{girth} of a graph is the length of its shortest cycle.). We provide a construction witnessing this fact in Theorem~\ref{thm:density-undirected}.

In a second part, we show that Theorem~\ref{thm:density-undirected} cannot be generalized to digraphs. Given a digraph $D$, the \textit{maximum average degree} of $D$ is defined as  $ \max\{  \frac{2|A(H)|}{|V(H)|}\mid \mbox{$H$ a non-empty subdigraph of $D$}\}$. The \textit{digirth} of a digraph is the length of its shortest directed cycle if the digraph contains cycles, and $+\infty$ otherwise.
In particular, oriented graphs are digraphs with digirth at least $3$.
It follows from Theorem~\ref{thm:min-degen} that every non $k$-mixing digraph $D$ contains a subdigraph $H$ with minimum out-degree and minimum in-degree at least $k-1$.
This shows that such a digraph $D$ has maximum average degree at least $2k-2$. This bound is tight because the bidirected complete digraph on $k$ vertices is $2k-2$-regular and is not $k$-mixing. However, unlike the undirected case, this is not the case for digraphs with arbitrary large digirth. In fact, this is not even the case for oriented graphs, which are exactly the digraphs with digirth at least 3. 
We pose the following conjecture:
\begin{conjecture}\label{conj:non2mixing}
    Any non 2-mixing oriented graph has maximum average degree at least 4.
\end{conjecture}
We prove two results providing some support for this conjecture. Firstly, using the Discharging Method, we prove the conjecture in the special case of freezable oriented graphs.

\begin{restatable}{theorem}{frozendensity}\label{theorem:2frozen-density}
    Let $\vec{G}=(V,A)$ be an oriented graph. If $\vec{G}$ is $2$-freezable, then $|A| \geq 2|V|$. 
\end{restatable}

From this result, we derive the following lower bound on the density of $k$-freezable oriented graphs.

\begin{restatable}{corollary}{kfrozen}\label{corollary:kfrozen-density}
     Let $\vec{G}=(V,A)$ be an oriented graph. If $\vec{G}$ is $k$-freezable, then  $|A| \geq k|V| + k(k-2)$. 
\end{restatable}
 We give a family of oriented graphs for which this bound is reached. Secondly, again with the Discharging Method, we show a statement weaker than Conjecture~\ref{conj:non2mixing} with $7/2$ instead of $4$.
 
\begin{restatable}{theorem}{notmixing}\label{theorem:not-2mixing-density}
Let $\vec{G}$ be an oriented graph.
If $\vec{G}$ is not 2-mixing, then $\Mad(\vec{G}) \geq \frac{7}{2}$. 
\end{restatable}

\medskip

Finally, in Section~\ref{section:open_problems},
we conclude with some open problems. 

\section{Complexity of  {\sc \texorpdfstring{$k$}{k}-Dicolouring Path}}
\label{section:complexity}

In this section, we establish some hardness results for {\sc $k$-Dicolouring Path}. We need some definitions. 

Given a graph $G=(V,E)$ together with a mapping $\phi : V \xrightarrow{} \{1,2\}$, an orientation $\vec{G}$ of $G$ is {\it proper} if for any $v\in V$, $d^-_{\vec{G}}(v) \geq \phi(v)$. A {\it reorienting sequence} from $\vec{G}_1$ to $\vec{G}_2$ is a sequence of proper orientations $\vec{\Gamma}_1,\dots,\vec{\Gamma}_p$ of $G$ such that $\vec{\Gamma}_1 = \vec{G}_1$, $\vec{\Gamma}_p = \vec{G}_2$, every $\vec{\Gamma}_i$ is proper and every $\vec{\Gamma}_{i+1}$ can be obtained from $\vec{\Gamma}_i$ by reversing exactly one arc.
The following problem has been shown to be PSPACE-complete in \cite{hearnTCS343} by a reduction from Quantified Boolean Formulas.

\defproblem{\sc Planar-Cubic-NCL}{ A cubic planar graph $G$, a mapping $\phi : V \xrightarrow{} \{1,2\}$, two proper orientations $\vec{G}_1$ and $\vec{G}_2$ of $G$.}{Is there a reorienting sequence from $\vec{G}_1$ to $\vec{G}_2$ ?}

\smallskip

We will derive the hardness results for  {\sc Dicolouring Path} from a hardness result on its list dicolouring version. 
Let $D$ be a digraph.
A \textit{list assignment} $L$ is a function which associates a list of colours to every vertex $v$ of $D$. An {\it $L$-dicolouring} of $D$ is a dicolouring $\alpha$ of $D$ such that $\alpha(v) \in L(v)$ for all vertex $v$. 
A {\it $k$-list assignment} is a list assignment $L$ such that $L(v)\subseteq [k]$ for all vertex $v$.
We denote by ${\cal D}(D,L)$ the redicolouring graph of the $L$-dicolourings of $D$. 
We call vertices $v$ such that $|L(v)|=1$ \emph{forced vertices}.
We will consider the following problem.

\defproblem{\sc $k$-List Dicolouring Path}
{A digraph $D$, a $k$-list assignment $L$, and two $L$-dicolourings $\alpha$ and $\beta$ of $D$.}
{Is there a path between $\alpha$ and $\beta$ in ${\cal D}(D,L)$ ?}

\smallskip

Let us start by proving the following result.

\begin{theorem}\label{thm:pspace-completeness-list}
    {\sc $2$-List Dicolouring Path} is PSPACE-complete on digraphs $D$ even when:
    \begin{itemize}
        \item forced vertices have degree at most $3$ and,
        \item either all the vertices have degree at most $5$ or, the digraph $D$ is planar and all the vertices have in and out-degree at most $3$.
    \end{itemize}
\end{theorem}

\begin{proof}
    First note that {\sc $2$-List Dicolouring Path} is indeed in NPSPACE. Given a digraph $D$ and two dicolourings $\alpha$ and $\beta$ of $D$ together with a redicolouring sequence from $\alpha$ to $\beta$, we can easily check with a polynomial amount of space that each dicolouring is valid. Then, we get that {\sc $k$-List Dicolouring Path} belongs to PSPACE thanks to Savitch's Theorem~\cite{savitchJCSS4}, which asserts that PSPACE = NPSPACE.

\medskip    
    
    We shall now give a polynomial reduction from \textsc{Planar-Cubic-NCL}.
    
    Let $G$ be a planar cubic graph on $n$ vertices $x_1,\dots,x_{n}$ with a mapping $\phi : V(G) \rightarrow \{1,2\}$. Let $\vec{G}_1$ and $\vec{G}_2$ be two proper orientations of $G$. 
    From $(G,\phi)$ we construct the digraph $D$ and the function $L$ as follows (see     Figure~\ref{fig-proof-thm-pspacecompleteness} for an illustration).
    \begin{itemize}
        \item For each vertex $x_i \in V(G)$, we create a {\it vertex-gadget} as follows. 
        
        We associate three vertices $x_{i,1},x_{i,2}$ and $x_{i,3}$ in $V_D$ so that each of these vertices is associated to exactly one edge of $G$ incident to $x_i$, and each edge of $G$ is associated to exactly two vertices of $D$. The function $L$ assigns to each of these vertices the list $\{1,2 \}$.
        
        We complete the vertex-gadget in two ways depending if $\phi(x_i) = 1$ or $2$. If $\phi(x_i) = 1$, then we create a new vertex $z_i$ in such a way that $(z_i, x_{i,1},x_{i,2},x_{i,3}, z_i)$ is a directed $4$-cycle. We set $L(z_i) = \{ 2 \}$.
        
        If $\phi(x_i)=2$, then we create three new vertices $z_{i,1},z_{i,2}$, and $z_{i,3}$ in such a way that $(x_{i,1},z_{i,1},x_{i,2},z_{i,2},x_{i,3},z_{i,3}, x_{i,1})$ is a directed $6$-cycle, and we add the arcs $x_{i,1}z_{i,2}$, $x_{i,2}z_{i,3}$ and $x_{i,3}z_{i,1}$. The function $L$ assigns the list $\{ 2 \}$ to $z_{i,1},z_{i,2}$, and $z_{i,3}$.
        
        \item For each edge $x_ix_j \in E(G)$, where $i<j$, we create a vertex $a_{ij}$ and create the directed $3$-cycle $(a_{ij}, x_{i,r}, x_{j,r'}, a_{ij})$, where $x_{i,r}$ and $x_{j,r'}$ are the vertices of $D$ corresponding to the edge $x_ix_j$ in $G$. We set $L(a_{i,j}) = \{ 1 \}$. This directed $3$-cycle is the {\it edge-gadget} of $x_ix_j$.
    \end{itemize}

\begin{figure}[hbtp]
  \begin{minipage}{\linewidth}
    \begin{center}	
      \begin{tikzpicture}[thick,scale=1, every node/.style={transform shape}]
	    \tikzset{vertex/.style = {circle,fill=black,minimum size=5pt,
                                        inner sep=0pt}}
	  	\tikzset{edge/.style = {->,> = latex'}}
	  
        \node[vertex, label=right:$x_i$, label=left:$1$] (xi) at  (0,1) {};
        \node[vertex, label=right:$x_j$, label=left:$2$] (xj) at  (0,-1) {};
        \draw[] (xi) to (xj);
        \draw[dashed] (xi) to (1.4,2.4);
        \draw[dashed] (xi) to (-1.4,2.4);
        \draw[dashed] (xj) to (1.4,-2.4);
        \draw[dashed] (xj) to (-1.4,-2.4);
        
        \node[vertex, label=right:{$x_{i,1}$~\scriptsize{$\{1,2\}$}}] (xi1) at  (8.4,2.4) {};
        \node[vertex, label=left:{{\scriptsize{$\{1,2\}$}}~$x_{i,2}$}] (xi2) at  (7,1) {};
        \node[vertex, label=left:{{\scriptsize{$\{1,2\}$}}~$x_{i,3}$}] (xi3) at  (5.6,2.4) {};
        \node[vertex, label=above:{$z_i$~\scriptsize{$\{2\}$}}] (zi) at  (7,2.4) {};
        
        \node[vertex, label=left:{{\scriptsize{$\{1,2\}$}}~$x_{j,1}$}] (xj1) at  (7,-1) {};
        \node[vertex, label=left:{{\scriptsize{$\{1,2\}$}}~$x_{j,2}$}] (xj2) at  (5.6,-3) {};
        \node[vertex, label=right:{$x_{j,3}$~\scriptsize{$\{1,2\}$}}] (xj3) at  (8.4,-3) {};
        \node[vertex, label=left:{{\scriptsize{$\{2\}$}}~$z_{j,1}$}] (zj1) at  (6.3,-2) {};
        \node[vertex, label={below:$z_{j,2}$~\scriptsize{$\{2\}$}}] (zj2) at  (7,-3) {};
        \node[vertex, label={right:$z_{j,3}$~\scriptsize{$\{2\}$}}] (zj3) at  (7.7,-2) {};
        \node[vertex, label={right:$a_{ij}$~\scriptsize{$\{1\}$}}] (aij) at  (7.8,0) {};
        
        \draw[edge] (xi1) to (xi2);
        \draw[edge] (xi2) to (xi3);
        \draw[edge] (xi3) to (zi);
        \draw[edge] (zi) to (xi1);
        \draw[edge, dashed] (xi3) to (4.2,3.8);
        \draw[edge, dashed] (5.6,3.8) to (xi3);
        \draw[edge, dashed] (xi1) to (9.8,3.8);
        \draw[edge, dashed] (8.4,3.8) to (xi1);
        
        \draw[edge] (xi2) to (xj1);
        \draw[edge] (xj1) to (aij);
        \draw[edge] (aij) to (xi2);
        
        \draw[edge] (xj1) to (zj1);
        \draw[edge] (zj1) to (xj2);
        \draw[edge] (xj2) to (zj2);
        \draw[edge] (zj2) to (xj3);
        \draw[edge] (xj3) to (zj3);
        \draw[edge] (zj3) to (xj1);
        
        \draw[edge] (xj1) to (zj2);
        \draw[edge] (xj3) to (zj1);
        \draw[edge] (xj2) to (zj3);
        
        \draw[edge, dashed] (xj3) to (9.8,-4.4);
        \draw[edge, dashed] (8.4,-4.4) to (xj3);
        \draw[edge, dashed] (xj2) to (4.2,-4.4);
        \draw[edge, dashed] (5.6,-4.4) to (xj2);
      \end{tikzpicture}
      \caption{An example of building $(D,L)$ from $(G,\phi)$, where $\phi(x_i) = 1$, $\phi(x_j) = 2$, and $i<j$.}
      \label{fig-proof-thm-pspacecompleteness}
    \end{center}    
  \end{minipage}
\end{figure}
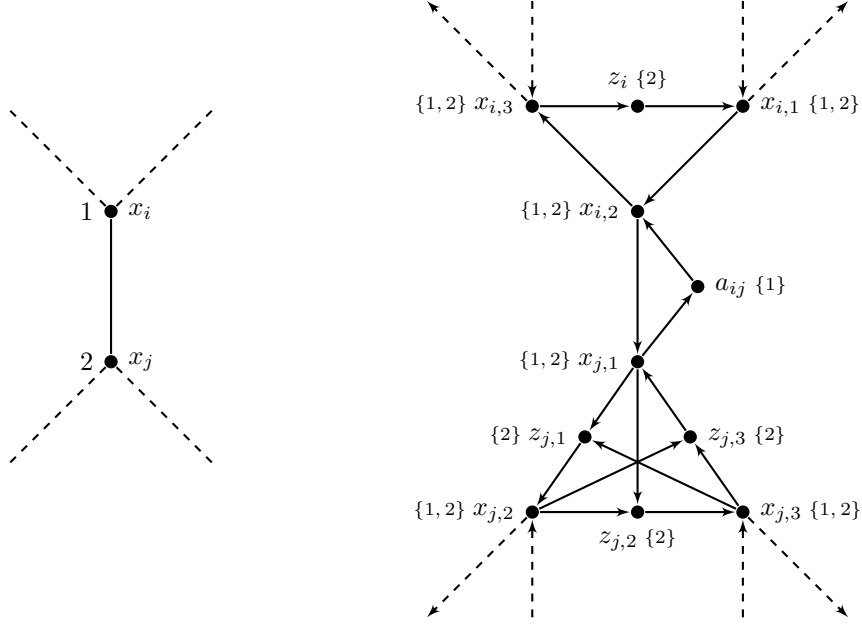

Note that the digraph $D$ has maximum degree at most $5$ and that its forced vertices have degree at most $3$.  

 \medskip
 
 For every proper orientation $\vec{G}$ of $G$, we define the \emph{dicolouring associated to $\vec{G}$}, a particular dicolouring of $D$ that we denote by $\alpha_{\vec{G}}$,  as follows.
    \begin{itemize}
        \item Forced vertices are assigned the colour of their list: vertices $z_i, z_{i,1},z_{i,2},z_{i,3}$ are coloured $2$, and vertices $a_{ij}$ are coloured $1$.
        \item 
        For each arc $x_ix_j\in A(\vec{G})$, we set $x_{i,r}$ to colour $2$ and $x_{j,r'}$ to colour $1$, where $x_{i,r}$ and $x_{j,r'}$ are the vertices of $D$ corresponding to the edge $x_ix_j$ in $G$.
    \end{itemize}
    \begin{claim}\label{clm:prop_valid_dicol}
    For every proper orientation $\vec{G}$ of $G$ and corresponding digraph $D$, the following hold.
    \begin{enumerate}
        \item[(i)] $\alpha_{\vec{G}}$ is an $L$-dicolouring of $D$.
        \item[(ii)] Unless $d^-_{\vec{G}}(x_i) = \phi(x_i)$ in $\vec{G}$, changing the colour of $x_{i,r}$ from $1$ to $2$ still yields a valid $L$- dicolouring of $D$. 
    \end{enumerate}
    \end{claim}
    \begin{proofclaim}
    Let us only show the first item, the second follows by similar arguments noting that $d^-_{\vec{G}}(x_i) > \phi(x_i)$ in this case.
    Note that by definition, $\alpha_{\vec{G}}$ satisfies the colouring constraints imposed by the list assignment $L$. 
    Let us then show that $\alpha_{\vec{G}}$ is indeed an $L$-dicolouring.
    Assume, for a contradiction, that there is a monochromatic directed cycle $C$ in $\alpha_{\vec{G}}$. For every edge-gadget, say corresponding to $x_i x_j$, the vertices $x_{i,r},x_{j,r'}$ must be coloured differently. Therefore, all vertices of $C$ must be contained in a single vertex-gadget of $D$.
    Let $x_i$ be the vertex such that $C$ is in the vertex-gadget of $x_i$.
    If $\phi(x_i)=1$, then $C$ must be $(z_i, x_{i,1},x_{i,2},x_{i,3}, z_i)$ and all its vertices should be coloured $2$.
    This is a contradiction since $d^-_{\vec{G}}(x_i)\geq 1$ implies, by construction, that at least one of $x_{i,1},x_{i,2},x_{i,3}$ is coloured~$1$.
    If $\phi(x_i)=2$, then $C$ contains at least two vertices in $\{x_{i,1},x_{i,2},x_{i,3}\}$ and two vertices in $\{z_{i,1},z_{i,2},z_{i,3}\}$.
    Then, at least two vertices of $\{x_{i,1},x_{i,2},x_{i,3}\}$ are coloured 2 because vertices $z_{i,j}$ are.
    This is a contradiction since $d^-_{\vec{G}}(x_i)\geq 2$ implies, by construction, that at least two vertices of $x_{i,1},x_{i,2},x_{i,3}$ are coloured $1$.
    \end{proofclaim}
    
    Let us take $\alpha_1 = \alpha_{\vec{G}_1}$ and $\alpha_2 = \alpha_{\vec{G}_2}$ to be the dicolourings obtained from the two proper orientations $\vec{G}_1,\vec{G}_2$.
    We will now show that there exists a reorienting sequence in $G$ from $\vec{G}_1$ to $\vec{G}_2$ if and only if there exists a redicolouring sequence in $D$ from $\alpha_1$ to $\alpha_2$.
    
    Assume first that there is a reorienting sequence $\vec{\Gamma}_1, \dots , \vec{\Gamma}_p$ from $\vec{G}_1$ to $\vec{G}_2$, and let us show how to build a corresponding redicolouring sequence.
    Consider any step $s$ of the reorienting sequence, say when $\vec{\Gamma}_s$ is transformed into $\vec{\Gamma}_{s+1}$ by reversing an arc $x_ix_j$ into $x_j x_i$.
    We will exhibit a path from $\alpha_{\vec{\Gamma}_s}$ to $\alpha_{\vec{\Gamma}_{s+1}}$ in ${\cal D}(D, L)$.
    Consider vertices $x_{i,r},x_{j,r'}$ in $D$, corresponding to the edge $x_ix_j$, and coloured $2$ and $1$ respectively in $\alpha_{\vec{\Gamma}_s}$.
    We first set the colour of $x_{j,r'}$ from $1$ to $2$.
    Since $a_{i,j}$ is forced to colour $1$, the edge-gadget is not monochromatic at this point.
    Moreover since step $s$ reorients arc $x_ix_j$ and still yields a proper orientation, $ d^-_{\vec{\Gamma}_s}(x_j) = d^-_{\vec{\Gamma}_{s+1}}(x_j) + 1 \geq  \phi(x_j)+1$. 
    The resulting colouring is an $L$-dicolouring by Claim~\ref{clm:prop_valid_dicol}~(ii).
    We then set the colour of $x_{i,r}$ from $2$ to $1$, yielding dicolouring $\alpha_{\vec{\Gamma}_{s+1}}$.
    Concatenating the redicolouring sequences obtained through the process above from steps $s=1$ to $s=p$ yields a redicolouring sequence from $\alpha_1$ to $\alpha_2$.
    
    \medskip
    
    Conversely, assume that there is a redicolouring sequence $\gamma_1, \dots , \gamma_p$ from $\alpha_1$ to $\alpha_2$.
    Observe that the only vertices of $D$ that are possibly recoloured in a step of our sequence are those defined as $x_{i,k}$ for $i \in [n]$ and $k \in [3]$, since all others are forced. 
    Now, at any step $s$ of the redicolouring, for each edge $x_ix_j$ of $G$, at most one of the two corresponding vertices is coloured 1, because $a_{ij}$ is forced to colour 1. 
    This allows us to define an orientation $\vec{\Gamma}_s$ of $G$ as follows.
    If the vertices $x_{i,r},x_{j,r'} \in V(D)$, corresponding to the gadget of edge $x_i x_j$, are not coloured the same in $\gamma_s$, orientation $\vec{\Gamma}_s$ sets $x_i x_j$ to be directed from the vertex coloured $2$ towards the vertex coloured $1$.
    Otherwise, both vertices are coloured $2$ in $\gamma_s$ and we preserve the orientation of the corresponding edge given by $\vec{\Gamma}_{s-1}$.
    In the first and last dicolourings, $\alpha_1$ and $\alpha_2$, for each edge $x_ix_j$ the corresponding vertices $x_{i,k}$ and $x_{j,k'}$ are coloured differently.
    Thus $\vec{\Gamma}_1=\vec{G}_1$ and $\vec{\Gamma}_p=\vec{G}_2$.
    Therefore, $\vec{\Gamma}_1, \dots , \vec{\Gamma}_p$ is a sequence of orientations of $G$ from $\vec{G}_1$ to $\vec{G}_2$ such that $\vec{\Gamma}_{s+1}$ is either obtained by reversing an arc of $\vec{\Gamma}_s$ (when one of the $x_{i,k}$ is recoloured to $1$, and the edge $x_ix_j$ whose edge-gadget contains $x_{i,k}$ was not oriented towards $x_i$), or equal to $\vec{\Gamma}_s$ otherwise  (and in particular when one of the $x_{i,k}$ is recoloured to $2$). 
    Moreover, at each step $s$, $\vec{\Gamma}_s$ is a proper orientation of $G$.
    Indeed, if $\phi(x_i)=1$ (resp. $\phi(x_i)=2$), then at least one vertex (resp. two vertices) of $\{x_{i,1},x_{i,2},x_{i,3}\}$ is coloured 1, and so $x_i$ has in-degree at least $1$ (resp. at least $2$) in $\vec{\Gamma}_s$. 
    Hence, taking the subsequence of $\vec{\Gamma}_1, \dots , \vec{\Gamma}_p$ that omits constant steps yields a reorienting sequence from $\vec{G}_1$ to $\vec{G}_2$.
    
    \medskip

    Since \textsc{Planar-Cubic-NCL} is PSPACE-complete, at this point our reduction already yields the PSPACE-completeness of \textsc{$2$-List Dicolouring Path}.
    By construction, forced vertices $z_{j,k}$ and $a_{ij}$ have degree at most $3$, and all other vertices have degree at most $5$. This achieves the proof of the first case of the result.
    
    Now, since the input instances are already planar, to get the PSPACE-completeness of \textsc{$2$-List Dicolouring Path} on planar digraphs, it suffices to use planar vertex and edge gadgets.
    In our current reduction, the only gadget which is not planar is the vertex-gadget corresponding to $x_i \in V(G)$ such that $\phi(x_i)=2$.
    We now consider the same reduction replacing the vertex-gadget for vertices such that $\phi(x_i)=2$ with a planar one. 
    For these vertices, the planar vertex-gadget is defined on the same set of vertices, i.e. $\{x_{i,1},x_{i,2},x_{i,3},z_{i,1},z_{i,2}, z_{i,3}\}$, but with the arcs of the directed 3-cycles $(z_{i,1}, x_{i,1}, x_{i,2}, z_{i,1})$, $(z_{i,3},x_{i,1},x_{i,3},z_{i,3})$ and $(z_{i,2},x_{i,3},x_{i,2},z_{i,2})$, as depicted in Figure~\ref{fig-planar-vertex-gadget}.
    This replacement produces a planar digraph in which all forced vertices still have degree at most $3$, and all vertices have maximum in- and out-degree at most $3$. This completes the proof.
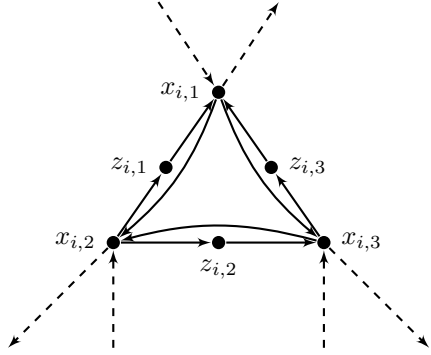
\begin{figure}[hbtp]
  \begin{minipage}{\linewidth}
    \begin{center}	
      \begin{tikzpicture}[thick,scale=1, every node/.style={transform shape}]
        \tikzset{vertex/.style = {circle,fill=black,minimum size=5pt,
                                        inner sep=0pt}}
        \tikzset{edge/.style = {->,> = latex'}}
        \node[vertex, label=left:$x_{i,1}$] (xi1) at  (7,-1) {};
        \node[vertex, label=left:$x_{i,2}$] (xi2) at  (5.6,-3) {};
        \node[vertex, label=right:$x_{i,3}$] (xi3) at  (8.4,-3) {};
        \node[vertex, label=left:$z_{i,1}$] (zi1) at  (6.3,-2) {};
        \node[vertex, label=below:$z_{i,2}$] (zi2) at  (7,-3) {};
        \node[vertex, label=right:$z_{i,3}$] (zi3) at  (7.7,-2) {};
        
        \draw[edge] (xi2) to (zi1);
        \draw[edge] (zi1) to (xi1);
        \draw[edge, bend left=15] (xi1) to (xi2);
        \draw[edge] (xi3) to (zi3);
        \draw[edge] (zi3) to (xi1);
        \draw[edge, bend right=15] (xi1) to (xi3);
        \draw[edge] (xi2) to (zi2);
        \draw[edge] (zi2) to (xi3);
        \draw[edge, bend right=15] (xi3) to (xi2);
        
        \draw[edge, dashed] (xi3) to (9.8,-4.4);
        \draw[edge, dashed] (8.4,-4.4) to (xi3);
        
        \draw[edge, dashed] (xi2) to (4.2,-4.4);
        \draw[edge, dashed] (5.6,-4.4) to (xi2);
        
        \draw[edge, dashed] (xi1) to (7.8,0.2);
        \draw[edge, dashed] (6.2,0.2) to (xi1);
      \end{tikzpicture}
      \caption{A planar vertex-gadget when $\phi(x_i)=2$}
      \label{fig-planar-vertex-gadget}
    \end{center}    
  \end{minipage}
\end{figure}

\end{proof}

The problem {\sc $k$-Colouring Path} is known to be PSPACE-complete for every $k\geq 4$ in the undirected case~\cite{bonsmaTCS410}.
Leveraging Theorem~\ref{thm:pspace-completeness-list}, we prove that this also holds for its dicolouring analogue for $k\geq 2$ colours, in both directed and oriented graphs.

\pspacecompleteness*

\begin{proof}
$(i)$
We give a reduction from {\sc $2$-List Dicolouring Path} on instances where forced vertices have degree at most $3$ and the graph has maximum degree $5$.
The problem is PSPACE-complete by Theorem~\ref{thm:pspace-completeness-list}.
Let $(D,L,\alpha_1,\alpha_2)$ be an instance of the problem, we construct an instance $(D',\alpha'_1,\alpha'_2)$ for {\sc $k$-Dicolouring Path} as follows.

We build $D'$ starting with $D'=D$. Then, for every vertex $v \in V(D)$, we let $\bid{K}_k^v$ be a bidirected complete graph on vertex set $\{z^v_i \mid i\in [k]\}$. We then add a digon between $v$ and each $z^v_i$ such that $i\notin L(v)$.
We define dicolourings $\alpha'_1$ and $\alpha'_2$ on $D'$ by extending dicolourings $\alpha_1$ and $\alpha_2$ as follows.
All vertices of $D'$ that were vertices of $D$ are coloured the same, and we set $z^v_i$ to colour $i$ for all $v\in V(D)$ and all $i\in [k]$.
Note that all the vertices of gadget $\bid{K}_k^v$ are then frozen in any $k$-dicolouring, letting us simulate in $D'$ the list dicolouring constraints on $D$.
An $L$-dicolouring path from $\alpha_1$ to $\alpha_2$ in $D$ is then exactly a dicolouring path from $\alpha'_1$ to $\alpha'_2$ in $D'$ restricted to vertices of $D$, achieving equivalence between the instances.

We will now show that the maximal degree of a vertex $u$ in $D'$ is $2k+1$.
If $u$ belongs to some gadget $\bid{K}_k^v$, then its degree is at most $2(k-1) + 2 = 2k$.
Note that when $u\in V(D)$, $u$ is of degree $2,3$ or $5$ in $D$, and our reduction adds exactly $2(k - |L(u)|)$ arcs incident to $u$.
If $|L(u)|=2$, this yields $d_{D'}(u) \leq 5 + 2k-4 = 2k+1$.
If $|L(u)|=1$, we know by construction that $d_D(u) \leq 3$, yielding $d_{D'}(u) \leq 3 + 2k-2 = 2k+1$.
This achieves the proof that $D'$ has maximum degree at most $2k+1$, concluding $(i)$.

\bigskip \noindent
$(ii)$
We give a reduction from {\sc $k$-Dicolouring Path} to {\sc $k$-Dicolouring Path} restricted to oriented graphs.
Let $(D,\alpha_1,\alpha_2)$ be an instance of {\sc $k$-Dicolouring Path}, we will build an equivalent instance $(\vec{G},\alpha'_1,\alpha'_2)$ where $\vec{G}$ is an oriented graph.
Take $\vec{H}$ to be an arbitrary oriented graph with dichromatic number exactly $k$.
We construct $\vec{G}$ from $D$ by replacing digons of $D$ as follows.
For each digon $[u,v]$ of $D$, create a copy $\vec{H}_{uv}$ of $\vec{H}$, then replace $[u,v]$ by a single arc from $u$ to $v$, and add all arcs from $v$ to $\vec{H}_{uv}$ and all arcs from $\vec{H}_{uv}$ to $u$.
By construction, $\vec{G}$ is an oriented graph.

In the following, we let $\xi$ be a fixed $k$-dicolouring of $\vec{H}$.
We show how to transform any $k$-dicolouring $\alpha$ of $D$ to a $k$-dicolouring $\alpha'$ of $\vec{G}$, and vice versa.
Given a $k$-dicolouring $\alpha$ for $D$, we define $\alpha'$ for $\vec{G}$ by colouring each copy $\vec{H}_{uv}$ of $\vec{H}$ with $\xi$, and keeping the same colours as $\alpha$ on $V(D)$.
Any monochromatic directed cycle in $(\vec{G},\alpha')$ must contain a vertex of some $\vec{H}_{uv}$, as otherwise it would be a subdigraph of $D$ and would already be monochromatic in $(D,\alpha)$.
Since $\xi$ is a dicolouring of $\vec{H}$, the cycle must contain both $u,v$, but then $u$ and $v$ being coloured the same would yield a monochromatic digon in $(D,\alpha)$, so $\alpha'$ is indeed a $k$-dicolouring of $\vec{G}$.  
Conversely, given any $k$-dicolouring $\alpha'$ of $\vec{G}$, we define $\alpha$ for $D$ as the restriction of $\alpha'$ on $V(D)$.
Similarly, if $(D,\alpha)$ were to contain a monochromatic directed cycle, any arc $(u,v)$ of the cycle that is not present in $\vec{G}$ may be replaced with $(u,w,v)$, taking $w \in \vec{H}_{uv}$ to be a vertex of the same colour as $u$ and $v$ (since $\dic(\vec{H}) = k$). This would yield a monochromatic directed cycle in $(\vec{G},\alpha')$, so $\alpha$ must be a $k$-dicolouring of $D$.

Now, we define the $k$-dicolourings $\alpha'_1,\alpha'_2$ on $\vec{G}$ obtained from $\alpha_1,\alpha_2$ by the transformation above, and let our output instance be $(\vec{G},\alpha'_1,\alpha'_2)$.
If there is a redicolouring sequence from $\alpha_1$ to $\alpha_2$ in $D$, we perform the same recolouring steps in $\vec{G}$ starting from $\alpha'_1$ and yielding $\alpha'_2$. Since we only recolour vertices of $V(D)$, the last paragraph yields that this sequence is valid.
Conversely, if there is a redicolouring sequence from $\alpha'_1$ to $\alpha'_2$, its restriction to $V(D)$ (omitting recolourings of vertices in subgraphs $\vec{H}_{uv}$) yields a valid sequence from $\alpha_1$ to $\alpha_2$ in $D$. This achieves the proof of the equivalence of the instances and proves $(ii)$.

\bigskip
\noindent
$(iii)$
We give a reduction from {\sc $2$-List Dicolouring Path} where $D$ is planar, forced vertices have degree at most $3$ and $D$ has in- and out-degree at most $3$.
The problem is PSPACE-complete by Theorem~\ref{thm:pspace-completeness-list}. Let $(D,L,\alpha_1,\alpha_2)$ be an instance of the problem. 
We make the same reduction as in the proof of $(i)$ by ensuring that $\bid{K}_k^v$ is embedded and coloured in such a way that (at most $3$) forbidden colours of $v$ lie on its external face.
This allows us to keep a planar representation of $D'$ which has maximum degree $2k+2$. This proves (iii).

\bigskip
\noindent
$(iv)$ As in $(iii)$, we give a reduction from {\sc $2$-List Dicolouring Path} where $D$ is planar, forced vertices have degree at most $3$ and every vertex has in- and out-degree at most $3$.
Since we are considering $2$-dicolourings, vertices with a list of size $2$ do not require a gadget to simulate forbidden colours.
The main difference with case $(iii)$ is that we cannot use a the bidirected complete graph $\vec{K}_2^v$ to freeze a vertex $v$ with list of size $1$.
To overcome this, we use the gadget depicted in Figure~\ref{fig:planar-frozen-vertex}, where the colour of all vertices is frozen.
Therefore, we simply have to attach such a gadget on each vertex with list size one, that is, those of the form $z_{i,r}$ or $a_{i,j}$. This can be done by creating a directed triangle including the vertex and two vertices of the opposite colour in the gadget, as depicted in the figure.

\begin{figure}[hbtp]
        \begin{minipage}{\linewidth}
            \begin{center}	
              \begin{tikzpicture}[thick,scale=1, every node/.style={transform shape}]
        	    \tikzset{vertex/.style = {circle,fill=black,minimum size=5pt,
                                    inner sep=0pt}}
                \tikzset{edge/.style = {->,> = latex'}}
        	    
                \node[vertex,red] (1r) at  (0,0) {};
                \node[vertex,red] (2r) at  (0,-1) {};
                \node[vertex,red] (3r) at  (0,-2) {};
                \node[vertex,red] (4r) at  (0,-3) {};
                \node[vertex,red] (5r) at  (0,-4) {};
                
                \node[vertex] (1b) at  (1,0) {};
                \node[vertex] (2b) at  (1,-1) {};
                \node[vertex] (3b) at  (1,-2) {};
                \node[vertex] (4b) at  (1,-3) {};
                \node[vertex] (5b) at  (1,-4) {};
                
                \node[vertex, label=left:$a$] (a) at  (-1,-1.5) {};
                
                \draw[edge] (1r) to (2r);
                \draw[edge] (2r) to (3r);
                \draw[edge] (3r) to (4r);
                \draw[edge] (4r) to (5r);
                
                \draw[edge] (1b) to (2b);
                \draw[edge] (2b) to (3b);
                \draw[edge] (3b) to (4b);
                \draw[edge] (4b) to (5b);
                
                \draw[edge] (1b) to (1r);
                \draw[edge] (2b) to (2r);
                \draw[edge] (3b) to (3r);
                \draw[edge] (4b) to (4r);
                \draw[edge] (5b) to (5r);
                
                \draw[edge] (2r) to (1b);
                \draw[edge] (3r) to (2b);
                \draw[edge] (4r) to (3b);
                \draw[edge] (5r) to (4b);
                
                \draw[edge] (a) to (2r);
                \draw[edge] (3r) to (a);
                
                \draw[] (1r) to[in=135, out=45] (1.3,0.3);
                \draw[edge] (1.3,0.3) to[in=45, out=-45] (2b);
                
                \draw[] (4r) to[in=135, out=-135] (-0.3,-4.3);
                \draw[edge] (-0.3,-4.3) to[in=-135, out=-45] (5b);
                
                \draw[] (5b) to[in=-45, out=60] (1.7,0.7);
                \draw[edge] (1.7,0.7) to[in=60, out=135] (1r);
              \end{tikzpicture}
          \caption{How to freeze the vertex $a$ in a planar oriented graph with two colours.}
          \label{fig:planar-frozen-vertex}
        \end{center}    
      \end{minipage}
    \end{figure}
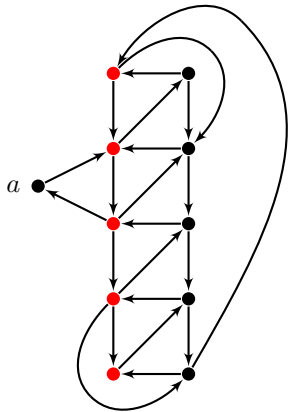
    
\end{proof}

\section{Connectivity and diameter of dicolouring graphs }
\label{section:diameter}

\subsection{Connectivity}

We start by showing some easy bounds on the minimal number of colours, with respects to variants of degeneracy, ensuring that digraphs or oriented graphs are $k$-mixing.
Bonsma and Cereceda~\cite{bonsmaTCS410}  and  Dyer et al.~\cite{dyerRSA29} independently proved that any (non-oriented) graph $G$ is $k$-mixing for every $k\geq \delta^*(G) +2$. Theorem~\ref{thm:min-degen}, which we recall, generalizes this result for digraphs.

\mindegen*

\begin{proof}
The proof is an adaption of the proof of~\cite{dyerRSA29} for undirected graphs.
We show the result by induction on the number of vertices of $D$, the result being obviously true for the digraph with one vertex. 
Let $D$ be a digraph on at least two vertices, $k\geq \delta^*_{\min}(D)+2$ be an integer and $\alpha,\beta$ be two $k$-dicolourings of $D$.
Let $v$ be a vertex such that $\min \{d^+(v),d^-(v)\} \leq \delta^*_{\min}(D)$ and let $D' = D - v$. 
By induction, $D'$ is $k$-mixing.
Let $\alpha',\beta'$ be the $k$-dicolourings of $D'$ induced by $\alpha,\beta$ (i.e. for every $v\in V(D')$, $\alpha'(v) = \alpha(v)$ and $\beta'(v) = \beta(v)$).
Since $D'$ is $k$-mixing, there exists a redicolouring sequence $\alpha'=\gamma_0,\dots,\gamma_\ell = \beta '$ from $\alpha'$ to $\beta'$, such that $\gamma_{i-1}$ and $\gamma_i$ differ in the colour of exactly one vertex of $\vec{G}'$ for $i \in  [1,\ell]$. We denote this vertex by $v_i$ and $\gamma_i(v_i)$ by $c_i$.

Let us prove that we can transform the redicolouring sequence from $D'$ into a redicolouring sequence for $D$. 
Starting from $\alpha$, we perform the same steps as in $\gamma_0,\dots,\gamma_\ell$ as long as they produce dicolourings of $D$.
If at some step $i$ in the sequence it is not possible to recolour vertex $v_i$ to $c_i$ in $D$, it must be because $v$ is currently coloured $c_i$ and recolouring $v_i$ to $c_i$ would create a monochromatic directed cycle containing both.
Assume that $v$ has at most $k-2$ out-neighbours (otherwise, $v$ has at most $k-2$ in-neighbours and the case is symmetrical).
Now, we can choose $c$ to be a colour different from $c_i$ that is also different from that of the out-neighbours of $v$.
Then, we recolour $v$ to $c$, allowing us to recolour $v_i$ to $c_i$ and continue the sequence.
\end{proof}

\avgdegen*

\begin{proof}
      We proceed by induction on the number of vertices of $\vec{G}$, the result being obviously true for the oriented graph with one vertex.  
      Consider now an oriented graph $\vec{G}$ on at least two vertices and $k\geq \left\lceil \delta^*_{\rm avg}(\vec{G}) \right\rceil +1$.
      Let $v$ be a vertex such that $d^+(v)+d^-(v) \leq 2\delta^*_{\rm avg}(\vec{G})$.
      By directional duality, we may assume $d^+(v) \leq d^-(v)$, then by assumption $d^+(v)\leq k-1$.
      We let $\vec{G}' = \vec{G} - v$.
      Then, $\delta^*_{\rm avg}(\vec{G}') \leq \delta^*_{\rm avg}(\vec{G})$, and the induction hypothesis yields that $\vec{G}'$ is $k$-mixing.
     
     Let $\alpha$ and $\beta$ be two $k$-dicolourings of $\vec{G}$, and let $\alpha',\beta'$ be the $k$-dicolourings of $\vec{G}'$ induced by $\alpha,\beta$.
     Since $\vec{G}'$ is $k$-mixing, there exists a sequence $\alpha'=\gamma_0,\dots,\gamma_\ell = \beta '$ of $k$-dicolourings of $\vec{G}'$ such that $\gamma_{i-1}$ and $\gamma_i$ differ in the colour of exactly one vertex of $\vec{G}'$ for $i \in  [1,\ell]$.
     We denote this vertex by $v_i$.
     Now we consider the same recolouring steps to recolour $\vec{G}$, starting from $\alpha$. If for some $i$ it is not possible to recolour $v_i$ to $c_i$, this must be because $v$ is currently coloured $c_i$ and recolouring $v_i$ to $c_i$ would create a monochromatic directed cycle. Note that such a cycle necessarily contains $v$, because $\gamma_i$ is a dicolouring of $G'$.
     \begin{itemize}
         \item If $d^+(v) \leq k-2$, then $v$ can be recoloured to $c$, a colour different from $c_i$ that does not appear on $N^+[v]$, allowing us to recolour $v_i$ to $c_i$ and proceed with the sequence.
         
         \item Otherwise $d^+(v) = k-1$. Then since $k\geq \left\lceil \delta^*_{\rm avg}(\vec{G}) \right\rceil +1$, we get $d^-(v) \leq k-1$, and then $d^-(v) = k-1$ (since $d^+(v) \le d^-(v)$).
          Since $\vec{G}$ is an oriented graph, and recolouring $v_i$ to $c_i$ would create a monochromatic directed cycle, then $v$ has at least one neighbour coloured $c_i$.
          Take $w$ to be one of these neighbours.
          Then, if $w\in N^+(v)$ (resp. $w\in N^-(v)$), we can recolour $v$ to a colour that does not appear in $N^+(v)$ (resp. $N^-(v)$), recolour $v_i$ to $c_i$ and continue the sequence.
     \end{itemize}
\end{proof}

Note that $\delta^*_{\max}(\vec{G})\geq \left\lceil \delta^*_{\rm avg}(\vec{G})\right\rceil$ since $\delta^*_{\max}(\vec{G})$ is an integer and $\delta^*_{\max}(\vec{G})\geq \delta^*_{\rm avg}(\vec{G})$. Consequently, every oriented graph $\vec{G}$ is $k$-mixing for all $k\geq \delta^*_{\max}(\vec{G})+1$.
A natural question is then whether this result can be extended to $\delta^*_{\rm out}$: is every oriented graph $\vec{G}$ $k$-mixing for all $k\geq \delta^*_{\rm out}(\vec{G})+1$?
This does not hold for directed graphs, as witnessed by the bidirected clique, and the following also answers the question in the negative for oriented graphs.

\begin{proposition}\label{prop:no-improv}
For every positive integer $k$, there exist $k$-out-degenerate oriented graphs that are not $(k+1)$-mixing.
\end{proposition}
\begin{proof}
Let $\vec{B}_0$ be the oriented graph with one vertex.
For $k\geq 1$, we construct $\vec{B}_{k}$ from $\vec{B}_{k-1}$ as follows: take two disjoint copies $\vec{B}_{k-1}^1$ and $\vec{B}_{k-1}^2$ of $\vec{B}_{k-1}$ and one vertex $r$; add all arcs from $r$ to $\vec{B}_{k-1}^1$, all arcs from $\vec{B}_{k-1}^1$ to $\vec{B}_{k-1}^2$ and all arcs from $\vec{B}_{k-1}^2$ to $r$. 
 
\begin{claim}\label{claim:Bk}
 $\vec{B}_k$ is $k$-out-degenerate and $\dic(\vec{B}_{k}) = k+1$. 
\end{claim}
 \begin{subproof}
 We prove the claim by induction on $k$, the result holding trivially for $k=0$.
 Assume now that $k \ge 1$, and consider $\vec{B}_k$.
 By induction, there exist two $(k-1)$-out-degeneracy orderings $\sigma_1$ and $\sigma_2$ on $\vec{B}^1_{k-1}$ and $\vec{B}^2_{k-1}$, respectively.
 In $\vec{B}_k$, each vertex of $V(\vec{B}_{k-1}^2)$ has exactly one out-neighbour in $V \setminus V(\vec{B}_{k-1}^2)$, namely $r$.
 We may then remove the vertices of $\vec{B}_{k-1}^2$ following $\sigma_2$, such that at each step the removed vertex has out-degree at most $k$.
 In the remaining graph, all out-neighbours of vertices in $V(\vec{B}_{k-1}^1)$ belong to $V(\vec{B}_{k-1}^1)$, allowing us to successively remove vertices of out-degree $k-1$ by following $\sigma_1$.
 Combining these facts, $\sigma_2 \cdot \sigma_1\cdot (r)$ yields a $k$-out-degeneracy ordering of $\vec{B}_{k}$.
 Hence $\vec{B}_{k}$ is $k$-out degenerate. 
 
The graph $\vec{B}_k$ is indeed $(k+1)$-dicolourable since any dicolouring such that the restriction to $\vec{B}^1_{k-1}$ and $\vec{B}^2_{k-1}$ is a $k$-dicolouring and $r$ is coloured with colour $k+1$ is a $(k+1)$-dicolouring of $\vec{B}_k$.
Now, assume for a contradiction that $\vec{B}_{k}$ has a $k$-dicolouring $\alpha$. Set $c = \alpha(r)$. By induction, $\dic(\vec{B}_{k-1}) = k$, so there is a vertex $x_1$ of $\vec{B}_{k-1}^1$ and a vertex $x_2$ of $\vec{B}_{k-1}^2$ such that $\alpha(x_1)= \alpha(x_2) = c$. But then $(r,x_1,x_2, r)$ is a monochromatic directed cycle, a contradiction. Thus $\dic(\vec{B}_{k}) = k+1$. 
\end{subproof}

Let $k$ be a positive integer.
Let $\vec{G}_k$ be the oriented graph obtained from the transitive tournament $TT_{k+1}$ on $k+1$ vertices, by adding, for each arc $xy$ of $TT_{k+1}$, a copy $\vec{B}_{k}^{xy}$ of $\vec{B}_k$, all arcs from $y$ to $\vec{B}_{k}^{xy}$ and all arcs from $\vec{B}_{k}^{xy}$ to $x$. Let us prove that $\vec{G}_k$ is $k$-out-degenerate and is not $(k+1)$-mixing.

\smallskip 

Let $(t_1,\dots, t_{k+1})$ be the {\it acyclic ordering} of $TT_{k+1}$ (such that there is no arc $t_it_j$ with $i > j$), and let $\sigma_{i,j}$ be a $k$-out-degeneracy ordering of $\vec{B}_{k}^{t_it_j}$ for all $1\leq i < j \leq k+1$.
We build a $k$-out-degeneracy ordering of $\vec{G}_k$ by combining these orders as follows: $(t_1) \cdot \sigma_{1,2} \cdot \sigma_{1,3} \cdot~\cdots~\cdot \sigma_{1,k+1} \cdot (t_2) \cdot \sigma_{2,3} \cdot \sigma_{2,4} \cdot~\cdots~\cdot \sigma_{2,k+1} \cdot (t_3) \cdot~\cdots~\cdot (t_k) \cdot \sigma_{k,k+1} \cdot (t_{k+1})$.
Thus $\vec{G}_k$ is $k$-out-degenerate.

\smallskip

We show that every $(k+1)$-dicolouring $\alpha$ of $\vec{G}_k$ is such that every vertex $v \in V(TT_{k+1})$ is frozen.
Since any $\alpha'$ defined from $\alpha$ by a permutation of the colours is also a $(k+1)$-dicolouring, the above being true yields that $\alpha'$ cannot be reached from $\alpha$ and implies that $\vec{G}_k$ is not $(k+1)$-mixing.
Consider $v \in V(TT_{k+1})$, and assume there is a redicolouring sequence starting from $\alpha$ which recolours $v$.
We let $\beta$ be the dicolouring of $\vec{G}_k$ right before $v$ is recoloured in the sequence, and $\beta'$ the dicolouring after recolouring $v$.
It suffices to note that $\beta(v) \neq \beta(w)$ for any $w \in V(TT_{k+1}) \backslash v$.
Indeed, since $\dic(B_{vw})=k+1$, there exists some $z \in V(B_{vw})$ coloured $\beta(v)$, and by construction $\beta(v) = \beta(w)$ would create a monochromatic directed cycle on vertices $v,w,z$.
Then, since $\beta'$ is obtained from $\beta$ by recolouring $v$, there exists some $w \in V(TT_{k+1})$ coloured the same as $v$ in $\beta'$, and the same argument yields a contradiction.

\end{proof}

\subsection{Diameter}

In this section, we prove Theorems~\ref{thm:linear-diameter}, \ref{thm:quadratic-diameter} and~\ref{thm:subcubic}.

\subcubic*

\begin{proof}
Let $\alpha$ and $\beta$ be any two $2$-dicolourings of $\vec{G}$. Let $x=\diff(\alpha,\beta)=|\{v \in V(\vec{G}) \mid \alpha(v) \neq \beta(v)\}|$. By induction on $x \geq 0$, let us show that there exists a path of length at most $2x$ from $\alpha$ to $\beta$ in ${\cal D}_2(\vec{G})$. This clearly holds for $x=0$ (i.e., $\alpha=\beta$). Assume $x>0$ and the result holds for every $x'<x$.

Let $v \in V(\vec{G})$ such that $1=\alpha(v) \neq \beta(v)=2$ ($v$ exists up to swapping the colours). If $v$ can be directly recoloured with colour $2$, then we recolour $v$ with colour $2$ and reach a new $2$-dicolouring $\alpha'$ such that $\diff(\alpha',\beta)=x-1$. Then, the result holds by induction.

Therefore, $v$ cannot be directly recoloured with colour $2$, i.e., there exists a directed cycle $C$ containing $v$ such that $\alpha(w)=2$ for every $w \in V(C) \setminus \{v\}$. Moreover, there must be a vertex $u \in V(C) \setminus \{v\}$ such that  $\beta(u)=1$. 

If $u$ is not a neighbour of $v$, then $u$ has two neighbours (those in $C$) coloured with $2$ in $\alpha$. Therefore, $u$ can be recoloured $1$ (it does not create any directed cycle of vertices coloured $1$ since $u$ has degree at most $3$ in $G$) and we reach a new $2$-dicolouring $\alpha'$ such that $\diff(\alpha',\beta)=x-1$. Then, the result holds by induction. Hence, $u$ is a neighbour of $v$ in $C$.

Again, we may assume that $u$ cannot be directly  recoloured $1$ (otherwise the result holds by induction) and so there exists a directed cycle $C'$ such that $\alpha(w)=1$ for every $w \in V(C') \setminus \{u\}$. Since $u$ and $v$ have degree at most $3$ in $\vec{G}$, and because of the $\alpha$-colours of vertices in $C$ and $C'$, then $V(C')\cap V(C) = \{v,u\}$. 
Let $h$ be the vertex of $V(C')\setminus \{u,v\}$ adjacent to $u$ in $C'$.
Since $\vec{G}$ is subcubic and because $h$ belongs to $C'$, $h$ cannot be in any directed cycle $C''$ such that all vertices of $C''$ but $h$ are coloured $2$ by $\alpha$. Hence, $h$ can be recoloured with $2$. In the obtained dicolouring, by choice of $h$, either the in- or the out-neighbourhood of $u$ (depending on the orientation of $uv$) is coloured $2$. Hence $u$ can be recoloured $1$, then $v$ can be recoloured with $2$ and $h$ recoloured with $1$, reaching a new $2$-dicolouring $\alpha'$ such that $\diff(\alpha',\beta)=x-2$, and  the result holds by induction.
\end{proof}

\lineardiam*

\begin{proof}
    The proof is very similar to the proof of Bousquet and Perarnau \cite{bousquetEJC52} for undirected graphs.
    Let $\alpha$ and $\beta$ be two $k$-dicolourings. Let us show by induction on the number of vertices that there exists a redicolouring sequence from $\alpha$ to $\beta$ where every vertex is recoloured at most $\delta_{\min}^*(D)+1$ times. 
    
    If $n = 1$ the result is obviously true. Let $D$ be a digraph on $n+1$ vertices, and let $u$ be a vertex such that $\min \{ d^+(u), d^-(u) \} \leq \delta_{\min}^*(D)$ and let $D' = D - u$.
    By directional duality, we may assume  $d^+(u) \leq \delta_{\min}^*(D)$.
    We denote by $\alpha'$ and $\beta'$ the dicolourings of $D'$ induced by $\alpha$ and $\beta$.
    By induction and since $\delta^*_{\min}(D') \leq \delta^*_{\min}(D)$, there exists a redicolouring sequence from $\alpha'$ to $\beta'$ such that each vertex is recoloured at most $\delta_{\min}^*(D)+1$ times.
    Now we consider the same recolouring steps to recolour $D$, starting from $\alpha$.
    If for some step $i$, it is not possible to recolour $v_i$ to $c_i$, this must be because $u$ is currently coloured $c_i$ and recolouring $v_i$ to $c_i$ would create a monochromatic directed cycle $C$.
    Since $u$ has at most $\delta_{\min}^*(D)$ out-neighbours, and since $k\geq 2\delta^*_{\min}(D)+2$, there are at least $\delta^*_{\min}(D)+2$ colours that do not appear in the out-neighbourhood of $u$.
    We choose $c$ among these colours so that $c$ does not appear in the next $\delta^*_{\min}(D)+1$ recolourings of $N^+(u)$ and we recolour $u$ with $c$. Note that $c$ is different from $c_i$ because either $v_i \in N^+(u)$ and $c_i$ appears in the next recolouring of $N^+(u)$ or the out-neighbour of $u$ in $C$ is currently coloured $c_i$. 
    
    Since $u$ has at most $\delta_{\min}^*(D)$ out-neighbours and since each vertex in $D'$ is recoloured at most $\delta_{\min}^*(D)+1$ times, there are at most $\delta_{\min}^*(D)(\delta_{\min}^*(D)+1)$ recolourings of an out-neighbour of $u$ in this redicolouring sequence. Hence, in this new redicolouring sequence, $u$ is recoloured at most $\delta_{\min}^*(D)$ times. We finally have to set $u$ to its colour in $\beta$.  Doing so $u$ is recoloured at most $\delta_{\min}^*(D)+1$ times. This concludes the proof.
\end{proof}

The goal of the rest of this part is to prove the following theorem which generalizes a result of Bousquet and Heinrich~\cite{bousquetHeinrichCoRR} to digraphs.

\quaddiam*

Given an undirected graph $G=(V,E)$, a list assignment $L$ is {\it $a$-feasible} if, for some ordering $v_1,\dots,v_n$ of $V$, $|L(v_i)| \geq |N(v_i) \cap \{v_{i+1},\dots,v_n\}| + 1 + a$ for every $i\in [n]$. 
 Bousquet and Heinrich proved the following result appearing as the first and second items of Theorem 6 in~\cite{bousquetHeinrichCoRR}:
\begin{theorem}[Bousquet and Heinrich~\cite{bousquetHeinrichCoRR}]
    \label{thm:bousquet-heinrich-2}
    Let $G$ be a graph and $a\in \mathbb{N}$. Let $L$ be an $a$-feasible list assignment and $k$ be the total number of colours. Then ${\cal C}(G,L)$ has diameter at most:
    \begin{itemize}
        \item[(i)] $kn$ if $k\leq 2a$,
        \item[(ii)] $Cn^2$ if $k\leq 3a$ (where $C$ a constant independent of $k,a$).
    \end{itemize}
\end{theorem}

For a graph $G$ and a list assignment $L$ of $G$, we say that an $L$-colouring $\alpha$ \emph{avoids a set of colours $S$} if for every vertex $v\in V(G)$, $\alpha(v)$ does not belong to $S$. In order to prove Theorem~\ref{thm:quadratic-diameter}, we deduce from Theorem~\ref{thm:bousquet-heinrich-2} the following lemma:

\begin{lemma}
    \label{lemma:avoid-colours}
    Let $G=(V,E)$ be an undirected graph on $n$ vertices, $k\in \mathbb{N}$ be the total number of colours, $L$ be a $\left \lceil \frac{k}{3} \right \rceil$-feasible list assignment and $\alpha$ an $L$-colouring of $G$ that avoids a set $S$ of $\left \lceil \frac{k}{3} \right \rceil$ colours. Then for any set of $\left \lceil \frac{k}{3} \right \rceil$ colours $S'$, there is an $L$-colouring $\beta$ of $G$ that avoids $S'$ and such that there is a recolouring sequence from $\alpha$ to $\beta$ of length at most $\frac{4k + 12}{3}n$.
\end{lemma}

The proof of Lemma~\ref{lemma:avoid-colours} comes from the proof of Lemma~8 in \cite{bousquetHeinrichCoRR}. We give it for the sake of completeness.

\begin{proof}
    Let $S'$ be any set of $\left \lceil \frac{k}{3} \right \rceil$ colours.
    
    We start with $G$ coloured by $\alpha$.
    Let $(v_1,\dots,v_n)$ be a degeneracy ordering of $G$. We consider each vertex from $v_n$ to $v_1$. For each vertex, if it possible we recolour it with a colour of $S$. We denote by $\eta$ the obtained $L$-colouring. This is done in less than $n$ steps. Observe that, for each colour $c \in S$ and each vertex $v_i$ of $G$, at least one of the following holds:
    \begin{itemize}
        \item $\eta(v_i) \in S$,
        \item $v_i$ has a neighbour in $\{v_{i+1},\dots,v_n\}$ coloured $c$, or
        \item $c\notin L(v_i)$.
    \end{itemize}
    
    Let $H$ be the subgraph of $G$ induced by the vertices whose colour in $\eta$ is not in $S$. We define $L_H$ by $L_H(v)= L(v)\setminus S$ for every $v\in V(H)$. Using the previous observation, we get that for every vertex $v_i$ of $H$, $v_i$ has at least $|L(v_i) \cap S|$ neighbours in $\{v_{i+1},\dots,v_n\} \setminus V(H)$. This implies that $L_H$ is a $\left \lceil \frac{k}{3} \right \rceil$-feasible list assignment of $H$ with a total number of colours bounded by $k-|S| \leq \frac{2k}{3}$. By Theorem~\ref{thm:bousquet-heinrich-2}~(i), the diameter of ${\cal C}(H,L_H)$ is at most $\frac{2k}{3}n$.
     Note that every recolouring of $H$ (valid for $L_H$) starting from $\eta_H$ (the colouring $\eta$ induced on $H$) gives a valid recolouring of $G$ starting from $\eta$. 
    
    Consider the following preference ordering on the colours: an arbitrary ordering of $[k]\setminus (S\cup S')$, followed by an ordering of $S'\setminus S$, and finally the colours from $S$. Let $\gamma$ be the $L$-colouring of $G$ obtained by colouring $G$ greedily from $v_n$ to $v_1$ with this preference ordering. Since $L$ is $\left \lceil \frac{k}{3} \right \rceil$-feasible, and $|S|=\left \lceil \frac{k}{3} \right \rceil$, no vertex is coloured with a colour in $S$ in $\gamma$. This implies that $\gamma_H$, the colouring $\gamma$ induced on $H$, is an $L_H$-colouring of $H$. Thus there is a recolouring sequence from $\eta_H$ to $\gamma_H$ of length at most $\frac{2k}{3}n$ steps. This gives a recolouring sequence in $G$. We can then recolour the vertices of $G - H$ to their target colour in $\gamma$ in at most $n$ steps. This shows that, in $G$, there is a recolouring sequence from $\alpha$ to $\gamma$ of length at most $n+ \frac{2k}{3}n + n = \frac{2k+6}{3}n$.
    
    Now observe that, for each colour $c \in R=[k] \setminus (S\cup S')$ and each vertex $v_i$ of $G$, at least one of the following must hold:
    \begin{itemize}
        \item $\gamma(v_i) \in R$,
        \item $v_i$ has a neighbour in $\{v_{i+1},\dots,v_n\}$ coloured $c$, or
        \item $c\notin L(v_i)$
    \end{itemize}
    Let $\Gamma$ be the subgraph of $G$ induced by all vertices coloured with a colour in $S'$ by $\gamma$. Note that $\Gamma$ is also the subgraph induced by all vertices coloured with a colour in $(S\cup S')$ by $\gamma$, because no vertex is coloured in $S$ by $\gamma$.  Let $L_\Gamma$ be the list assignment defined by $L_{\Gamma}(v) = L(v)\cap (S\cup S')$ for all $v\in V(\Gamma)$. By the previous observation, $L_\Gamma$ is $\left \lceil \frac{k}{3} \right \rceil$-feasible, and the total number of colours is $|S\cup S'| \leq 2\left \lceil \frac{k}{3} \right \rceil$. Thus by Theorem~\ref{thm:bousquet-heinrich-2}~(i), ${\cal C}(\Gamma,L_\Gamma)$ has diameter at most $2\left \lceil \frac{k}{3} \right \rceil n$. Let $\beta_\Gamma$ be an $L_\Gamma$-colouring of $\Gamma$ that avoids the colours of $S'$ (such a colouring exists because $|S'|= \left \lceil \frac{k}{3} \right \rceil$ and $L_\Gamma$ is $\left \lceil \frac{k}{3} \right \rceil$-feasible) and $\gamma_\Gamma$ the colouring $\gamma$ induced on $\Gamma$.
    
    There is a recolouring sequence of length at most $2\left \lceil \frac{k}{3} \right \rceil n$ from $\gamma_\Gamma$ to $\beta_\Gamma$. This extends to a recolouring sequence in $G$ from $\gamma$ to $\beta$ where $\beta$ does not use any colour of $S'$.
    
    The total number of steps to reach $\beta$ from $\alpha$ is at most $\frac{2k+6}{3}n + 2\left \lceil \frac{k}{3} \right \rceil n $ which is bounded by $\frac{4k + 12}{3}n$. This shows the result.
\end{proof}

\addtocounter{theorem}{-8}
We are now able to prove Theorem~\ref{thm:quadratic-diameter}.
\begin{proof}[Proof of Theorem~\ref{thm:quadratic-diameter}]
    Let $D=(V,A)$ be a digraph on $n$ vertices and $k\geq \frac{3}{2}(\delta^*_{\min}(D)+1)$. 
    Let $(v_1,\dots,v_n)$ be a min-degeneracy-ordering of $D$, that is an ordering such that for each $i\in [n]$, $v_i$ has at most $\delta^*_{\min}(D)$ out-neighbours or at most $\delta^*_{\min}(D)$ in-neighbours in $\{v_{i+1},\dots,v_n\}$.
    
    We define $B$ a subset of $A$ as follows: for each $i\in [n]$, if $v_i$ has at most $\delta^*_{\min}(D)$ out-neighbours in $\{v_{i+1},\dots,v_n\}$, we add all arcs of $A$ from $v_i$ to $\{v_{i+1},\dots,v_n\}$ to $B$. Otherwise, we add all arcs of $A$ from $\{v_{i+1},\dots,v_n\}$ to $v_i$ to $B$.
    Note that both digraphs $(V,B)$ and $(V,A\setminus B)$ are acyclic.
    
    Let $H$ be $(V,B)$ and $G$ be the underlying graph of $H$. Note that, since $(V,A\setminus B)$ is acyclic, each colouring of $G$ is a dicolouring of $D$, but a dicolouring of $D$ is not necessarily a colouring of $G$.
    
    By construction, $G$ has degeneracy at most $\delta^*_{\min}(D)$. Using Theorem~\ref{thm:bousquet-heinrich}, we get that ${\cal C}_k(G)$ has diameter at most $C_0n^2$ for some constant $C_0$ independent from $k$.
    
    \medskip
     
     Set $\delta^*=\delta^*_{\min}(D) \geq \delta^*(G)$, $X_i=\{v_{i+1},\dots,v_n\}$, and $H_i = G-X_i$ for all $i\in [n]$.
     
    Let $\alpha$ be any dicolouring of $D$.
    Let $L_i$ be the list assignment of $H_i$ defined by
    $$L_i(v_j) = [k] \setminus \{ \alpha(v) \mid v\in N_G(v_j) \cap X_i \}~\mbox{for~all~} j\in [i].$$
      Since $k$, the total number of colours, is at least $\frac{3}{2}(\delta^* + 1)$, for every $j\in [i]$ we have:
    \begin{align*}
        |L_i(v_j)| &\geq k - |N_G(v_j) \cap X_i|\\
                    &\geq \frac{k}{3} + \frac{2}{3}\frac{3}{2}(\delta^* +1) - |N_G(v_j)\cap X_i|\\
                    &\geq |N_G(v_j)\cap \{v_{j+1},\dots,v_i\}| + 1 + \frac{k}{3}
    \end{align*}
    Hence, since $|L_i(v_j)|$ is an integer, $L_i$ is a $\left \lceil \frac{k}{3} \right \rceil$-feasible list assignment of $H_i$.
    
    \begin{remark}\label{remark:extendstoD}
    Let $\gamma$ be a dicolouring of $D$ and for some $i$, $\gamma$ agrees with $\alpha$ on $\{v_{i+1},\dots,v_n\}$ and $\gamma_{H_i}$ is an $L_i$-colouring of $H_i$. Then any recolouring sequence starting from $\gamma_{H_i}$ on $H_i$, valid for $L_i$, is a valid redicolouring sequence in $D$. Assume this is not the case and at one step, we get to an $L_i$-colouring $\zeta$ of $H_i$ but $\zeta_D$ contains a monochromatic directed cycle $C$, where $\zeta_D(v) = \zeta(v)$ when $v$ belongs to $H_i$ and $\zeta_D(v) = \gamma(v)$ otherwise. Let $v_j$ be the vertex of $C$ such that $j$ is minimum. This vertex $v_j$ has both an in-neighbour $v_{j_1}$ and an out-neighbour $v_{j_2}$ coloured $\zeta_D(v_j)$ such that $j_1,j_2 \geq j$. We know that either $v_{j_1}v_j$ or $v_jv_{j_2}$ belongs to $G$. Assume by symmetry that $v_{j_1}v_j$ belongs to $G$. Then either $j_1\leq i$ and then $v_{j_1}v_j$ is a monochromatic edge in $H_i$ or $j_1\geq i+1$ but then $\zeta(v_{j_1}) = \alpha(v_{j_1})$ does not belong to $L_i(v_j)$. In both cases, we get a contradiction.
    \end{remark}

    \addtocounter{theorem}{-3}
    
    \begin{claim}\label{claim+k/3}
    Let $\gamma_i$ be a dicolouring of $D$ which induces an $L_i$-colouring of $H_i$ avoiding at least $\left \lceil \frac{k}{3} \right \rceil$ colours in $H_i$. 
There is a redicolouring sequence of length at most $\frac{8k + 24}{3}n + \left \lceil \frac{k}{3} \right \rceil$ from $\gamma_i$ to a dicolouring $\gamma_{i+\left \lceil \frac{k}{3} \right \rceil}$ which induces an $L_{i+\left \lceil \frac{k}{3} \right \rceil}$-colouring of $H_{i+\left \lceil \frac{k}{3} \right \rceil}$ avoiding at least $\left \lceil \frac{k}{3} \right \rceil$ colours in $H_{i+\left \lceil \frac{k}{3} \right \rceil}$.
    \end{claim}
    
    \addtocounter{theorem}{+3}
    \begin{subproof}
    Let $S$ be a set of colours of size exactly $\left \lceil \frac{k}{3} \right \rceil$ avoided by $\gamma_i$ on $H_i$.
            For each vertex $v_j$ in $\{v_{i+1},\dots,v_{i+\left \lceil \frac{k}{3} \right \rceil} \}$, we choose a colour $c_j$ so that each of the following holds:
    \begin{itemize}
        \item $c_j$ belongs to $L_j(v_j)$,
        \item $c_j$ does not belong to $\{ \alpha(u) \mid u\in N_G(v_j) \cap \{v_{j+1},\dots,v_n\} \}$,
        \item for each $\ell\in \{i+1,\dots,j-1\}$, $c_\ell$ is different from $c_j$.
    \end{itemize}
    Note that this is possible because $L_j(v_j)$ is $\left \lceil \frac{k}{3} \right \rceil$-feasible.
    Now let $S'$ be $\{c_{i+1},\dots, c_{i+\lceil\frac{k}{3}\rceil}\}$. Observe that $|S'| = \left \lceil \frac{k}{3} \right \rceil$. By Lemma~\ref{lemma:avoid-colours}, there is, in $H_i$, a recolouring sequence  of length at most $\frac{4k+12}{3}n$, valid for $L_i$, from $\gamma_i$ to some $\gamma_i'$ that avoids $S'$. This recolouring sequence extends to a redicolouring sequence in $D$ by Remark~\ref{remark:extendstoD}.
    In the obtained dicolouring, since $\gamma_i'$ avoids $S'$ on $H_i$, we can recolour successively $v_j$ with $c_j$ for all $i+1 \leq j \leq i+\left \lceil \frac{k}{3} \right \rceil$. This does not create any monochromatic directed cycle by choice of $c_j$. Let $\eta_i$ be the resulting dicolouring of $D$. Now we define $\Tilde{L}_i$ a list-assignment of $H_i$ as follows: 
    
    $$\Tilde{L}_i(v_j) = [k] \setminus \{ \eta_i(v) \mid v\in N(v_j) \cap \{v_{i+1},\dots,v_n\}))$$
    
    Using the same arguments as we did for $L_i$, we get that $\Tilde{L}_i$ is $\left \lceil \frac{k}{3} \right \rceil$-feasible for $H_i$. Note that $\eta_i$ is an $\Tilde{L}_i$-colouring of $H_i$ that avoids $S'$. Let $S''$ be a set of $\left \lceil \frac{k}{3} \right \rceil$ colours disjoint from $S'$. By Lemma~\ref{lemma:avoid-colours}, there is, in $H_i$, a recolouring sequence (valid for $\Tilde{L}_i$) of length at most $\frac{4k+12}{3}n$ from $\eta_i$ to some $\eta_i'$ that avoids $S''$. This recolouring sequence extends directly to a redicolouring sequence in $D$. Since $S'$ is disjoint from $S''$, the obtained dicolouring is an $L_{i+\left \lceil \frac{k}{3} \right \rceil}$-colouring of $H_{i+\left \lceil \frac{k}{3} \right \rceil}$ that avoids at least $\left \lceil \frac{k}{3} \right \rceil$ colours in $H_{i+\left \lceil \frac{k}{3} \right \rceil}$. Hence we get a redicolouring sequence from $\gamma_i$ to the desired $\gamma_{i+\left \lceil \frac{k}{3} \right \rceil}$, in at most $\frac{8k + 24}{3}n + \left \lceil \frac{k}{3} \right \rceil$ steps. This proves Claim~\ref{claim+k/3}.
    \end{subproof}
    
     Note that $\gamma_{\left \lceil \frac{k}{3} \right \rceil}$ can be reached from $\alpha$ in less than $n$ steps: for all $j\in [\left \lceil \frac{k}{3} \right \rceil]$, choose a colour $c_j$ so that each of the following holds:
    \begin{itemize}
        \item $c_j$ belongs to $L_j(v_j)$,
        \item $c_j$ does not belong to $\{ \alpha(u) \mid u\in N_G(v_j) \cap \{v_{j+1},\dots,v_n\} \}$,
        \item for each $\ell\in [j-1]$, $c_\ell$ is different from $c_j$.
    \end{itemize} 
    Now we can recolour successively $v_1,\dots,v_{\left \lceil \frac{k}{3} \right \rceil}$ to their corresponding colour in $\{c_1,\dots,c_{\left \lceil \frac{k}{3} \right \rceil} \}$. Then applying Claim~\ref{claim+k/3} iteratively at most $\left \lfloor \frac{n}{\left \lceil \frac{k}{3} \right \rceil} \right \rfloor \leq \frac{3n}{k}$ times, there is a redicolouring sequence of length at most $n + \frac{3n}{k}\left(\frac{8k + 24}{3}n + \left \lceil\frac{k}{3} \right \rceil \right)$ from $\alpha$ to a dicolouring of $D$ that is also a colouring of $G$.
     Note that there exists a constant $C_1$ such that
     $n + \frac{3n}{k}\left(\frac{8k + 24}{3}n + \left \lceil\frac{k}{3} \right \rceil \right) \leq C_1n^2$.
     
 \medskip
 
 Let $\alpha$ and $\beta$ be two $k$-dicolourings of $D$.
 As proved above, there is a redicolouring sequence of length at most $C_1n^2$ from $\alpha$ (resp. $\beta$) to a dicolouring $\alpha'$ (resp. $\beta'$) of $D$ that is also a colouring of $G$.
Since  ${\cal C}_k(G)$ has diameter at most $C_0n^2$, there is 
a recolouring sequence of $G$ of length at most $C_0n^2$ from $\alpha'$ to $\beta'$, which is also a redicolouring sequence of $D$ (since every colouring of $G$ is a dicolouring of $D$).
The union of those three sequences yields a redicolouring sequence from $\alpha$ to $\beta$ of length at most 
    $(2C_1+C_0) n^2$.
\end{proof}

\addtocounter{theorem}{8}

\section{Density of non 2-mixing oriented graphs}

\label{section:density}
\subsection{Density of non \texorpdfstring{$k$}{k}-mixing undirected graphs}

Let $k\in \mathbb{N}$. As observed in the introduction, every non $k$-mixing graph has maximum average degree at least $k-1$. This bound is tight because the complete graph on $k$ vertices is $(k-1)$-regular and is not $k$-mixing. Moreover, it is shown in~\cite{bonamyENDM68} that this bound is tight even when we restrict to graphs of arbitrary large girth. The initial proof uses the probabilistic method. In this section, we give a new constructive proof of this result, based on an explicit construction of regular bipartite graphs from Lazebnik and Ustimenko in~\cite{lazebnikDAM60}.

\begin{theorem}[Bonamy, Bousquet and Perarnau,~\cite{bonamyENDM68}]
    \label{thm:density-undirected}
    For any $k,\ell \in \mathbb{N^*}$, there exists a $(k-1)$-regular $k$-freezable graph $G_{k,\ell}$ with girth at least $\ell$.
\end{theorem}

We first make the following remark that we will use in the proof of Theorem~\ref{thm:density-undirected}:
\begin{remark}\label{rem:color-class}
    Let $k\in \mathbb{N}^*$, $G$ be a $(k-1)$-regular $k$-freezable graph and $\alpha$ be a frozen $k$-colouring of $G$, then all colour classes of $\alpha$ have the same size. This follows from the fact that, for every vertex $v$ of $G$, $N[v]$ use all colours in $\alpha$. Thus, given two colours $i,j \in [k]$, there must be a perfect matching in $G$ between the vertices coloured $i$ and the vertices coloured $j$. In particular, this implies that the number of vertices coloured $i$ is the same as the number of vertices coloured $j$.
\end{remark}

\begin{proof}[Proof of Theorem~\ref{thm:density-undirected}]
    Let us fix $\ell \in \mathbb{N}$. 
    We prove the statement by induction on $k$, the result holding trivially for $k = 1$. Let $k > 1$ and assume that there exists a $(k-2)$-regular $(k-1)$-freezable graph $G_{k-1,\ell}$ with girth at least $\ell$. Let $\alpha$ be a frozen $k$-colouring of $G_{k-1,\ell}$.
    
    We denote by $n$ the number of vertices of $G_{k-1,\ell}$. Consider $H$ an $n$-regular bipartite graph with girth at least $\ell$ (such a graph exists by a construction from Lazebnik and Ustimenko~\cite{lazebnikDAM60}). Since $H$ is bipartite, we can colour the edges of $H$ with exactly $n$ colours such that two adjacent edges receive different colours.
    By Remark~\ref{rem:color-class}, all colour classes of $\alpha$ have the same size. Thus there is an ordering $(v_1,\dots,v_n)$ of $V(G_{k-1,\ell})$ such that for each $i\in [n-k+1]$, the vertices $v_i,\dots,v_{i+k-1}$ have different colours in $\alpha$.  
    
    We denote by $(A,B)$ the bipartition of $H$. 
    Let $G_{k,\ell}$ be the graph obtained from $H$ as follows.
    \begin{itemize}
    \item For each $a \in A$, replace $a$ by a copy $G^a$ of $G_{k-1,\ell}$, and connect $v^a_i$ (the vertex corresponding to $v_i$ in $G^a$) to the edge coloured $i$ that was incident to $a$. 
    
    \item For each $b\in B$, replace $b$ by an independent set $I^b=\{x^b_1,\dots,x^b_{\frac{n}{k}}\}$ of size $\frac{n}{k}$ (by Remark~\ref{rem:color-class}, $\frac{n}{k}$ is an integer). Connect $x^b_i$ to the edges coloured $\{ k(i-1)+1,\dots,ki \}$ that were incident to $b$. 
    \end{itemize}
    
    Observe that $G_{k,\ell}$ is $k$-regular: every vertex in a $G^a$ is adjacent to its $k-1$ neighbours in $G^a$ and exactly one neighbour in one of the $I^b$s;  every vertex in an $I^b$ has exactly $k$ neighbours by construction.
    Moreover, $G_{k,\ell}$ has girth at least $\ell$. Indeed, assume, for a contradiction, that it contains a cycle $C$ of length at most $\ell - 1$.
    Then $C$ cannot contain an edge of $H$, otherwise, contracting each copy of $G^a$ would transform $C$ into a cycle of length at most $\ell-1$ in $H$. Thus $C$ must be contained in some $G^a$, which is a copy of $G_{k-1,\ell}$, which is impossible since $G_{k-1,\ell}$ has girth at least $\ell$.
    
    Let $\beta$ be the $(k+1)$-colouring of $G_{k,\ell}$ such that the restriction of $\beta$ to each $G^a$ corresponds to $\alpha$, and $\beta(x^b_i) = k+1$ for all $b\in B$ and $i\in [n/k]$.
     Let $v$ be a vertex of $G$.
     If $v$ belongs to some $G^a$, then since $\alpha$ is frozen in $G_{k-1,\ell}$, $N_{G^a}[v]$ contains all colours of $[k]$. Moreover, by construction, $v$ has a neighbour in some $I^b$ which is coloured $k+1$. 
     If $v$ is in some $I^b$, then $v$ is coloured $k+1$ and by construction it has exactly one neighbour in each colour class.
     In both cases, $N[v] = [k+1]$. Thus no vertex can be recoloured and so $\beta$ is a frozen colouring of $G_{k,\ell}$.
\end{proof}

\subsection{Density of non \texorpdfstring{$2$}{2}-mixing and \texorpdfstring{$2$}{2}-freezable oriented graphs}

Let $k\in \mathbb{N}$ and $D$ be a digraph that is not $k$-mixing. As observed in the introduction,  Theorem~\ref{thm:min-degen} implies that $\Mad(D) \geq 2k-2$. This bound is tight because the bidirected complete digraph on $k$ vertices is $(2k-2)$-regular and is not $k$-mixing. However, unlike the undirected case, this result does not extend to digraphs with larger digirth, even for $k=2$. 
While the above inequality states that every non $2$-mixing digraph has maximum average degree at least $2$, we conjecture
that every non $2$-mixing oriented graph has maximum average degree at least $4$.

\begin{conjecture}
    \label{conjecture-non-mixing}
    Any non $2$-mixing oriented graph has maximum average degree at least $4$.
\end{conjecture}

\begin{remark}\label{rem:F}
If true, this conjecture would be tight since there exist $2$-freezable oriented graphs with maximum average degree $4$.
Consider for example the oriented graph $\vec{F}_n$ obtained from the disjoint union of two disjoints directed paths $(u_1, \dots , u_n)$ and $(v_1, \dots , v_n)$ by adding the set of arcs
$\{u_iv_i \mid i\in [n]\} \cup \{v_{i+1}u_i \mid i\in [n-1]\} \cup \{v_1u_2, u_nv_1, v_{n-1}u_n\}$ (see Figure~\ref{fig:2frozen-density-tightness} for an illustration).
Let $\alpha$ be the $2$-dicolouring of $\vec{F}_n$ in which all the $u_i$ are coloured $1$ and all the $v_i$ are coloured $2$.
One can easily check that $\Mad(\vec{F}_n)=4$ and $\alpha$ is a $2$-frozen dicolouring of $\vec{F}_n$. 
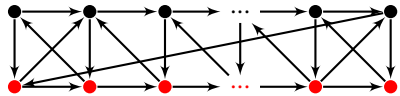
\begin{figure}[hbtp]
    \begin{minipage}{\linewidth}
        \begin{center}	
          \begin{tikzpicture}[thick,scale=1, every node/.style={transform shape}]
    	    \tikzset{vertex/.style = {circle,fill=black,minimum size=5pt,
                                inner sep=0pt}}
            \tikzset{edge/.style = {->,> = latex'}}
    	   \begin{scope} [rotate=90]
            \node[vertex,red] (1r) at  (0,0) {};
            \node[vertex,red] (2r) at  (0,-1) {};
            \node[vertex,red] (3r) at  (0,-2) {};
            \node[vertex,red] (5r) at  (0,-4) {};
            \node[vertex,red] (6r) at  (0,-5) {};
            
            \node[vertex] (1b) at  (1,0) {};
            \node[vertex] (2b) at  (1,-1) {};
            \node[vertex] (3b) at  (1,-2) {};
            \node[vertex] (5b) at  (1,-4) {};
            \node[vertex] (6b) at  (1,-5) {};
            
            \end{scope}
            
            \node[] (4b) at  (3,1) {...};
            \node[red] (4r) at  (3,0) {...};
            \draw[edge] (1r) to (2r);
            \draw[edge] (2r) to (3r);
            \draw[edge] (3r) to (4r);
            \draw[edge] (4r) to (5r);
            \draw[edge] (5r) to (6r);
            
            \draw[edge] (1b) to (2b);
            \draw[edge] (2b) to (3b);
            \draw[edge] (3b) to (4b);
            \draw[edge] (4b) to (5b);
            \draw[edge] (5b) to (6b);
            
            \draw[edge] (1b) to (1r);
            \draw[edge] (2b) to (2r);
            \draw[edge] (3b) to (3r);
            \draw[edge] (4b) to (4r);
            \draw[edge] (5b) to (5r);
            \draw[edge] (6b) to (6r);
            
            \draw[edge] (2r) to (1b);
            \draw[edge] (3r) to (2b);
            \draw[edge] (4r) to (3b);
            \draw[edge] (5r) to (4b);
            \draw[edge] (6r) to (5b);
            
            \draw[edge] (6b) to (1r);
            \draw[edge] (5r) to (6b);
            \draw[edge] (1r) to (2b);
          \end{tikzpicture}
      \caption{The 2-freezable oriented graph $\vec{F}_n$ and a frozen $2$-dicolouring.}
      \label{fig:2frozen-density-tightness}
    \end{center}    
  \end{minipage}
\end{figure}
\end{remark}

We now prove two results supporting Conjecture~\ref{conjecture-non-mixing}. First we prove that Conjecture~\ref{conjecture-non-mixing} holds with the stronger assumption that $G$ is $2$-freezable.

\frozendensity*

\begin{proof}
     Let $\vec{G}=(V,A)$ be a $2$-freezable oriented graph, and $\alpha$ a frozen $2$-dicolouring of $\vec{G}$.
     For a vertex $v \in V$, we say that a vertex $u \in V$ is \textit{blocking} for $v$ (in dicolouring $\alpha$), if one of the following holds:
     \begin{itemize}
         \item $u$ is an out-neighbour of $v$, $\alpha(u) \neq \alpha(v)$, and there exists a directed path $(u,...,x,v)$ such that $(u,...,x)$ is monochromatic, or
         
         \item $u$ is an in-neighbour of $v$, $\alpha(u) \neq \alpha(v)$, and there exists a directed path $(x,...,u,v)$ such that $(x,...,u)$ is monochromatic.
     \end{itemize} 

     We shall use a discharging argument. 
     We set the initial charge of every vertex $v$ to be $d(v) = d^+(v) + d^-(v)$. Observe that $d(v) \geq 2$ otherwise $v$ can be recoloured in $\alpha$.
     We then use the following discharging rule.
     \begin{itemize}
     \item[(R)] every vertex receives 1 from each of its blocking neighbours.
     \end{itemize}
     Let $f(v)$ be the final charge of every vertex $v$.
     Let us show that $f(v) \geq 4$ for every $v \in V$.

     Let $v \in V$. We assume without loss of generality that $\alpha(v) =1$. Since $\alpha$ is frozen, $v$ admits at least one out-neighbour $v^+$ and one in-neighbour $v^-$ coloured $2$ that are blocking, and thus sending $1$ to $v$ by (R).
     Let us now examine the charge that $v$ sends to others vertices. Let $w$ be a vertex to which $v$ sends charge. The vertex $v$ is blocking for $w$, so $\alpha(w)=2$.
     Moreover if $w$ is an out-neighbour (resp. in-neighbour) of $v$, then $v$ has an in-neighbour (resp. out-neighbour) coloured $1$.
     We are in one of the following cases.
     \begin{itemize}
         \item If $v$ sends no charge, then $f(v) \geq d(v) +2 \geq 4$.
         \item If $v$ sends charge only to some out-neighbours, then it does not send to its in-neighbours. Since $v$ has at least two in-neighbours (one blocking $v$ and one coloured $1$), $f(v) \geq d(v) + 2 - (d(v) -2) \geq 4$.
         \item If $v$ sends charge only to some in-neighbours, symmetrically to above, $f(v)\geq 4$.
         \item If $v$ sends charge only to both out-neighbours and in-neighbours, then its has both an in-neighbour and an out-neighbour coloured $1$ to which it sends no charge. Hence $f(v) \geq d(v) + 2 - (d(v) -2) \geq 4$.
     \end{itemize}
In all cases, we have $f(v)\geq 4$.
Consequently, $2|A| = \sum_{v\in V} d(v) = \sum_{v\in V} f(v) \geq 4|V|$.
\end{proof}

We can deduce from Theorem~\ref{theorem:2frozen-density} the following lower bound on the density of a $k$-freezable oriented graph.

\kfrozen*

\begin{proof}
    Suppose for a contradiction that there is a $k$-freezable oriented graph $\vec{G}=(V,A)$ such that $|A| < k|V| + k(k-2)$.
    Without loss of generality, we may take $\vec{G}$ having a minimum number of arcs among all such graphs.
    Let $\alpha$ be a frozen $k$-dicolouring of $\vec{G}$.
    For each $i,j\in [k]$, let $\vec{G}_i$ be the subdigraph of $\vec{G}$ induced by the vertices coloured $i$ in $\alpha$, and let $\vec{G}_{i,j}$ be the subdigraph of $\vec{G}$ induced by the vertices coloured $i$ or $j$ in $\alpha$. We set $n_i=|V(\vec{G}_i)|$, $m_i=|A(\vec{G}_i)|$ and $m_{i,j}=|A(\vec{G}_{i,j})|$.
    
    We first show that, for any  $i\in [k]$, $m_i \leq n_i -1$. Suppose not. Then, since $\vec{G}_i$ is acyclic, it admits an acyclic ordering $(x_1,\dots, x_{n_i})$.
    Now consider $\vec{G}' = (\vec{G} \setminus A(\vec{G}_i)) \cup \{x_jx_{j+1} \mid j \in [n_i-1] \}$ with the same dicolouring $\alpha$.
    Clearly $|A(\vec{G}')| < |A(\vec{G})|$.
    Let $v$ be a vertex of $\vec{G}'$.
    If $v\in V(\vec{G}_i)$, then $x$ is still blocked in $(\vec{G}',\alpha)$ because it is blocked in $(\vec{G},\alpha)$. 
    Now, suppose $v\notin V(\vec{G}_i)$.
    For any colour $j$ distinct from $i$ and $\alpha(v)$, it is impossible to recolour $v$ with $j$ because it was already impossible in $\vec{G}$. 
    Now, in $\vec{G}$, it was impossible to recolour $v$ to $i$, so there is a directed path in $\vec{G}_i$ whose initial vertex $x_k$ is an out-neighbour of $v$ and whose terminal vertex $x_{\ell}$ is an in-neighbour of $v$. Since $(x_1,\dots, x_{n_i})$ is an acyclic ordering of $\vec{G}_i$, we have $k\leq \ell$.
    Thus $(x_k, \dots , x_{\ell})$ is a directed path in $\vec{G}'$.
    Hence $v$ cannot be recoloured to $i$ in $(\vec{G}',\alpha)$, meaning it is also blocked in $(\vec{G}',\alpha)$.
    Since all vertices of $(\vec{G}',\alpha)$ are blocked, $\alpha$ is a frozen $k$-dicolouring of $\vec{G}'$, contradicting the minimality of $\vec{G}$.
    
    We will now prove the result by bounding $S = \sum_{1\leq i < j \leq k} m_{i,j}$.
    First, since $\alpha$ induces a 2-frozen dicolouring on $\vec{G}_{i,j}$, Theorem~\ref{theorem:2frozen-density} yields $m_{i,j} \geq 2(n_i + n_j)$ for any $1\leq i < j \leq k$.
    Thus, 
    \[
        S = \sum_{1\leq i < j \leq k} m(\vec{G}_{i,j}) \geq 2\sum_{1\leq i < j \leq k} (n_i + n_j) = 2n(k-1)\ .
    \]
    Next, observe that $S = |A(\vec{G})| + \sum_{i=1}^k(k-2)m_i$ because every monochromatic arc of $D$ is counted exactly $(k-1)$ times in $S$, and every other arc only once.
    Thus
    \begin{align*}
        S &= |A(\vec{G})| +  (k-2)\sum_{i=1}^km_i\\
            &< kn + k(k-2) + (k-2)\sum_{i=1}^k(n_i-1)\\
            &= 2n(k-1)\ .
    \end{align*}
    Putting the two inequalities together, we get $2n(k-1) \leq S < 2n(k-1)$, which is a contradiction.
\end{proof}

\begin{remark}
    The bound of Corollary~\ref{corollary:kfrozen-density} is tight: we can extend the construction of $\vec{F}_n$ (defined in Remark~\ref{rem:F}) to $k$-freezable oriented graphs $\vec{F}_n^k$ with exactly $k|V(\vec{F}_n^k)|+k(k-2)$ arcs. The oriented graph $\vec{F}_n^k$  is constructed as follows. We start from $k$ disjoint directed paths $P_1,\dots,P_k$ of length $n$. For each $j\in [k]$, let $P_s=(v^j_1,\dots,v^j_n)$.  For each pair $1\leq j<\ell \leq k$, we add the set of arcs $\{v^j_iv^\ell_i \mid i\in [n]\} \cup \{v^\ell_{i+1}v^j_i \mid i\in [n-1]\} \cup \{v^\ell_1v^j_2, v^j_nv^\ell_1, v^\ell_{n-1}v^j_n\}$, so that the subdigraph induced by $V(P_j) \cup V(P_{\ell})$ is isomorphic to $\vec{F}_n$. By construction, $|A(\vec{F}_n^k)|=k|V(\vec{F}_n^k)|+k(k-2)$.
    
    Let $\alpha_k$ be the dicolouring of $\vec{F}_n^k$ assigning colour $j$ to the vertices of $P_j$ for all $j\in [k]$. 
    Since $V(P_j) \cup V(P_{\ell})$ is isomorphic to $\vec{F}_n$, then every vertex of $P_j$ cannot be recoloured with $\ell$ (and vice versa). Therefore $\alpha_k$ is a frozen $k$-dicolouring of $\vec{F}_n^k$. 
\end{remark}

\notmixing*

\addtocounter{theorem}{-10}

\begin{proof}
    Let $\vec{G}=(V,A)$ be an oriented graph which is not 2-mixing, and take $\vec{G}$ to be minimal for this property (every proper induced subdigraph $\vec{H}$ of $\vec{G}$ is 2-mixing). 
    In order to get a contradiction, assume that $|A| < \frac{7}{4}|V|$.
    Let $\alpha$ and $\beta$ be two 2-dicolourings of $\vec{G}$ such that there is no redicolouring sequence from $\alpha$ to $\beta$.
    We will first prove some structural properties on $\vec{G}$, which we then leverage through a discharging strategy to show that $|A| \geq \frac{7}{4}|V|$.
    
    \begin{claim} \label{claim:nosource-nosink}
        $\vec{G}$ has no source nor sink.
    \end{claim}
    \begin{proofclaim}
        Assume that $\vec{G}$ contains a vertex $s$ which is either a source or a sink. 
        By minimality of $\vec{G}$, we know that $\vec{G}' = \vec{G} - s$ is 2-mixing.
        In particular, there is a redicolouring sequence $\gamma_0',\dots,\gamma_r'$ where $\gamma_0'$ and $\gamma_r'$ are the restrictions of $\alpha$ and $\beta$ to $\vec{G}'$ respectively.
        
        For all $i\in [r]$, let $\gamma_i$ be defined by $\gamma_i(s) = \alpha(s)$ and $\gamma_i(v) = \gamma_i'(v)$ for all $v\neq s$.
        Since $s$ is either a source or a sink, $\gamma_i$ is a $2$-dicolouring of $\vec{G}$. 
        If $\beta(s) = \alpha(s)$, then $\beta =\gamma_r$ and so $\gamma_0,\dots,\gamma_r$ is a redicolouring sequence from $\alpha$ to $\beta$, a contradiction. Thus $\beta(s) \neq \alpha(s)$. Then setting $\gamma_{r+1}=\beta$, we have that $\gamma_0,\dots,\gamma_r,\gamma_{r+1}$ is a redicolouring sequence from $\alpha$ to $\beta$, a contradiction. 
    \end{proofclaim}
    
    \begin{claim}\label{claim:delta3}
        $\delta(\vec{G}) = 3$.
    \end{claim}
    
    \begin{proofclaim}
        First note that if $\delta(\vec{G}) \geq 4$ then $|A| \geq 2|V|$, contradicting our assumption that $|A| < \frac{7}{4}|V|$. Therefore $\delta(\vec{G}) \leq 3$.
    
        Assume now that $\vec{G}$ contains a vertex $u$ such that $d(u) \leq 2$.
        By Claim~\ref{claim:nosource-nosink} we know that $d^+(u) = d^-(u) = 1$.
        Let $\vec{G}'$ be $\vec{G} - u$.
        By minimality of $\vec{G}$, we know that $\vec{G}'$ is 2-mixing.
        Let $\gamma_0',\dots,\gamma_r'$ be a redicolouring sequence in $\vec{G}'$ where $\gamma_0'$ and $\gamma_r'$ are the restrictions of $\alpha$ and $\beta$ to $\vec{G}'$. 
        Towards the contradiction, we exhibit a redicolouring sequence from $\alpha$ to $\beta$.
        To do so, we show, for any $i \in \{0, 1, \dots ,r\}$, the existence of a $2$-dicolouring $\gamma_i$ of $\vec{G}$ such that $\gamma'_i$ is the restriction of $\gamma_i$ on $\vec{G}'$, and there is a redicolouring sequence from $\alpha$ to $\gamma_i$.
        
        For $i=0$, the result holds trivially with $\gamma_0 =\alpha$.
        Assume now that $\gamma_{i-1}$ exists, and that there exists a redicolouring sequence reaching $\gamma_{i-1}$ from $\alpha$.
        Let $x_i$ be the vertex such that $\gamma'_{i-1}(x_i) \neq \gamma'_{i}(x_i)$.
        Consider the operation of recolouring $x_i$ by its opposite colour in $\gamma_{i-1}$.
        If this creates no monochromatic directed cycle, we let $\gamma_i$ be the output dicolouring.
        Otherwise, this creates a monochromatic directed cycle containing $u$ and its two neighbours $u^-$ and $u^+$.
        Since in $\gamma'_{i-1}$, the colour of $x_i$ is different from at least one of $\{u^-, u^+\}$, we may first recolour $u$ and then $x_i$ to obtain the desired $\gamma_i$.
        At the end of this process, we obtain a redicolouring sequence from $\alpha$ to a 2-dicolouring $\gamma_r$ agreeing with $\beta$ on $V(\vec{G}')$.
        If necessary, we may then recolour $u$ to yield a redicolouring sequence from $\alpha$ to $\beta$, achieving the contradiction and yielding $\delta(\vec{G})=3$.
    \end{proofclaim}

    For every positive integer $i$, an {\it $i$-vertex} (resp. {\it $(\geq i)$-vertex}) is a vertex of degree $i$ (resp. at least $i$) in $\vec{G}$, and the set of $i$-vertices in $\vec{G}$ is denoted $V_i$.
    Let $u$ be a $3$-vertex.
    By Claim~\ref{claim:nosource-nosink}, either $d^-(u) =1$ or $d^+(u) =1$.
    Then, the {\it lonely neighbour} of $u$ is its unique in-neighbour if $d^-(u) =1$ and its unique out-neighbour if $d^+(u) =1$.  Observe that, in any dicolouring, recolouring $u$ with a colour different from the one of its lonely neighbour yields another dicolouring, because every directed cycle containing $u$ must contain its lonely neighbour. We will use this argument several times along the remaining of this proof.

    \begin{claim}\label{claim:existence_of_phi_and_psi}
        Let $u$ be a 3-vertex and $v$ its lonely neighbour. There exists two $2$-dicolourings $\phi$ and $\psi$ that agree on $V(\vec{G}) \setminus \{u,v\}$ and such that there is no redicolouring sequence from $\phi$ to $\psi$.
    \end{claim}
    
    \begin{proofclaim}
        Assume for a contradiction that for any pair $\phi,\psi$ of $2$-dicolourings that agree on $V(\vec{G}) \setminus \{u,v\}$, there is a redicolouring sequence from $\phi$ to $\psi$. 
        Let $\vec{G}' = \vec{G} - u$, and $\alpha',\beta'$ be the restrictions of our initial dicolourings $\alpha, \beta$ to $\vec{G}'$.
        By minimality of $\vec{G}$, there is a redicolouring sequence $\alpha'=\gamma_0',\dots,\gamma_r'=\beta'$ in $\vec{G}'$.
        In order to get a contradiction, we will extend the latter into a redicolouring sequence from $\alpha$ to $\beta$.
        As in Claim~\ref{claim:delta3}, for $i=0$ to $r$, we show inductively that there exists a $2$-dicolouring $\gamma_i$ of $\vec{G}$ such that $\gamma'_i$ is the restriction of $\gamma_i$ on $\vec{G}'$, and there is a redicolouring sequence from $\alpha$ to $\gamma_i$.
        
        For $i=0$ the result holds trivially with $\gamma_0 =\alpha$.
        Assume now that $\gamma_{i-1}$ exists, and let $x_i \in V(\vec{G}')$ be the vertex recoloured at step $i$, such that $\gamma'_{i-1}(x_i) \neq \gamma'_{i}(x_i)$. We assume without loss of generality that $\gamma'_{i}(x_i) = 1$.
        We try to recolour $x_i$ to colour $1$ in $\gamma_{i-1}$.
        If this creates no monochromatic directed cycle, then we have the desired $\gamma_i$.
        If it does, the fact that $\gamma'_i$ is a dicolouring of $\vec{G}'$ yields that any resulting monochromatic directed cycle must contain both $u$ and its lonely neighbour $v$. This implies that $\gamma_{i-1}(u) = 1$, and we distinguish two cases, depending on the colour of $v$ in $\gamma_{i-1}$.
        \begin{itemize}
            \item If $\gamma_{i-1}(v)=1$, then $\gamma_{i-1}(v) \neq \gamma_{i-1}(x_i) = 2$, which implies that $x_i$ and $v$ are different.
            In this case, we can first recolour $u$ and then $x_i$ to obtain the desired $\gamma_i$ from $\gamma_{i-1}$, which combined with the sequence from $\alpha$ to $\gamma_{i-1}$ obtained by induction yields the redicolouring sequence from $\alpha$ to $\gamma_i$.
            
            \item Else, $\gamma_{i-1}(v) = 2$ and since recolouring $x_i$ creates a monochromatic directed cycle containing both $u$ and $v$, it must be because $x_i=v$.
            Now, let $\gamma_i$ be the $2$-dicolouring of $\vec{G}$ obtained from $\gamma_{i-1}$ by swapping the colours of $u$ and $v$ (\textit{i.e.} $\gamma_i(u) = 2$ and $\gamma_i(v) = 1$). Observe that $\gamma_i$ is a $2$-dicolouring since $\gamma_i'$ is, $v$ is the lonely neighbour of $u$, and $\gamma_i(u) \neq \gamma_i(v)$.
            Since $\gamma_i$ and $\gamma_{i-1}$ agree on $V(\vec{G}) \setminus \{u,v\}$, our assumption yields a redicolouring sequence from $\gamma_{i-1}$ to $\gamma_{i}$.
            Then, our induction hypothesis yields a sequence from $\alpha$ to $\gamma_{i-1}$, which,combined with the one above, yields a sequence from $\alpha$ to $\gamma_{i}$.
        \end{itemize} 
        Therefore, by induction, we obtain a redicolouring sequence from $\alpha$ to a 2-dicolouring $\gamma_r$ which agrees with $\beta$ on $V(\vec{G}')$.
        Then, either $\gamma_r=\beta$, or we may recolour $u$ in $\gamma_r$ to obtain $\beta$, which in any case yields a redicolouring sequence from $\alpha$ to $\beta$, a contradiction.
    \end{proofclaim}

    \begin{claim}\label{claim:degree_at_least_4}
        Let $u$ be a 3-vertex and $v$ its lonely neighbour. 
        Then $d^+(v) \geq 2$ and $ d^-(v) \geq 2$.
    \end{claim}
    
    \begin{proofclaim}
        By directional duality, we may assume $d^+(u) = 1$, such that $uv$ is an arc of $\vec{G}$. 
        
        \medskip
        
        Let $\phi$ and $\psi$ be two $2$-dicolourings of $\vec{G}$ given by Claim~\ref{claim:existence_of_phi_and_psi}.
        That is, both dicolourings agree on $V(\vec{G}) \setminus \{ u,v \}$, and there is no redicolouring sequence from $\phi$ to $\psi$.
        Without loss of generality, we assume that $\phi(u) = 1$. Note that $\phi$ and $\psi$ must differ on both vertices $u,v$, for otherwise $\phi, \psi$ would yield a trivial sequence from $\phi$ to $\psi$. Moreover, $u$ cannot be recoloured in $\phi$, for otherwise we can successively recolour $u$ and $v$ to obtain a redicolouring sequence between $\phi$ and $\psi$. Altogether, this shows that $\psi(u) = \phi(v) = 2$ and $\phi(u) = \psi(v) = 1$.
        Finally, there must be two paths from $v$ to $u$ such that one is internally coloured $1$, and the other is internally coloured $2$, respectively, in both $\phi$ and $\psi$ (see Figure~\ref{fig:claim_164}).
        Otherwise, either $v$, respectively $u$, could be recoloured in $\phi$ or $\psi$ to get a $2$-dicolouring in which $u$ and $v$ have the same colour allowing us to get a contradiction as above.
        This shows $d^+(v)\geq 2$.

        \begin{figure}[H]
            \begin{minipage}{\linewidth}
                \begin{center}	
                  \begin{tikzpicture}[thick,scale=1, every node/.style={transform shape}]
            	    \tikzset{vertex/.style = {circle,fill=black,minimum size=5pt,
                                        inner sep=0pt}}
                    \tikzset{edge/.style = {->,> = latex'}}
            	    
                    \node[vertex,red, label=left:$u$] (u) at  (-1,0) {};
                    \node[red] (q) at  (0,1) {$\cdots$};
                    
                    \node[vertex,blue, label=right:$v$] (v) at  (1,0) {};
                    \node[blue] (q2) at  (0,-1) {$\cdots$};
                    
                    \draw[edge] (u) to (v);
                    
                    \draw[edge,blue] (v) to[out=-90, in=0] (q2);
                    \draw[edge,blue] (q2) to[out=180, in=-90] (u);

                    \draw[edge,red] (v) to[out=90, in=0] (q);
                    \draw[edge,red] (q) to[out=180, in=90] (u);
                  \end{tikzpicture}
              \caption{A $3$-vertex $u$ and its lonely neighbour $v$ in a dicolouring $\phi$ from which we cannot reach the dicolouring $\psi$ swapping the colours of $u$ and $v$ via a redicolouring sequence.
              Vertices coloured $1$ are shown in red, while vertices coloured $2$ are shown in blue.} 
              \label{fig:claim_164}
            \end{center}    
          \end{minipage}
        \end{figure}
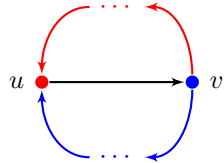
        
        \medskip
        
        Now, assume for the sake of contradiction that $d^-(v) = 1$, and recall that $\alpha,\beta$ are two $2$-dicolourings witnessing that $\vec{G}$ is non-mixing.
        Up to recolouring $u$ or $v$ in $\alpha$ or $\beta$ if both vertices are coloured the same, we may assume $\alpha(u) \neq \alpha(v)$ and $\beta(u) \neq \beta(v)$. We assume without loss of generality $\beta(u)=1$ (implying $\beta(v) = 2$).
        We first show that we can further assume $\alpha(u) = 1$ (and therefore $\alpha(v) = 2$).
        To do so, if $\alpha(u) = 2$, we build a redicolouring sequence from $\alpha$ to another $2$-dicolouring $\tilde{\alpha}$ such that $\tilde{\alpha}(u) = 1$.
        Let $t_1$ and $t_2$ be the two in-neighbours of $u$.
        If $t_1$ and $t_2$ are coloured the same in $\alpha$, then we can either recolour $u$ then $v$ if $\alpha(t_1) =\alpha(t_2) =2$, or recolour $v$ then $u$ if $\alpha(t_1) =\alpha(t_2) =1$, yielding the desired $\tilde{\alpha}$.
        Hence we may assume that $\alpha(t_1)=1$ and $\alpha(t_2) = 2$.
        Let $\vec{G}' =\vec{G} - \{u,v\}$ and $\alpha'$ be the restriction of $\alpha$ to $\vec{G}'$. 
        Given a $2$-dicolouring $\zeta$ of $\vec{G}$, we define its {\it mirror} $\bar{\zeta}$ as the $2$-dicolouring of $\vec{G}$ such that $\zeta(x) \neq \bar{\zeta}(x)$ for all $x\in V(\vec{G})$.
        By minimality of $\vec{G}$, there is a redicolouring sequence from $\alpha'$ to its mirror $\bar{\alpha'}$ in $\vec{G}'$.
        Since the colours of all vertices are swapped, there must be a $2$-dicolouring $\gamma'$ at some point in the sequence where $\gamma'(t_1) = \gamma'(t_2)$ (for instance the first time $t_1$ or $t_2$ is recoloured).
        We extend the sequence from $\alpha'$ to $\gamma'$ into a redicolouring sequence from $\alpha$ to $\gamma$ in $\vec{G}$, where $u$ and $v$ are constantly coloured $2$ and $1$ in both dicolourings. This is possible because $v$ (respectively $u$) is the lonely neighbour of $u$ (respectively $v$).
        Now, since $\gamma(t_1) = \gamma(t_2)$, in $\gamma$ we may exchange the colours of $u$ and $v$ as above to obtain the desired $\tilde{\alpha}$.
        
        Since $\alpha$ can be obtained from $\tilde{\alpha}$ through a redicolouring sequence, we can redefine $\alpha$ to be $\tilde{\alpha}$.
        Now, $\alpha(u) = \beta(u) = 1$ and $\alpha(v) = \beta(v)=2$, and we claim there exists a redicolouring sequence from $\alpha$ to $\beta$.
        Indeed, let $\alpha'$ and $\beta'$ be the the restrictions of $\alpha$ and $\beta$ to $\vec{G}'$.
        By minimality of $\vec{G}$, there is a redicolouring sequence from $\alpha'$ to $\beta'$.
        This sequence extends to $\vec{G}$ by fixing the colour of $u$ and $v$ to $1$ and $2$ respectively. This gives a contradiction, and yields $d^-(v) \geq 2$.
    \end{proofclaim}
    
    \begin{claim}\label{claim:v_has_degree5}
        Let $u$ be a 3-vertex such that $|N(u) \cap V_3| = 2$ and $v$ its lonely neighbour, then:
        \begin{enumerate}[(i)]
            \item $d(v) \geq 5$, and
            \item $d(v) = 5$ only if $v$ is adjacent to two $(\geq 4)$-vertices.
        \end{enumerate}
    \end{claim}
    
    \begin{proofclaim}
        By directional duality, we may assume $d^+(u) = 1$.
        Let $t_1$ and $t_2$ be the in-neighbours of $u$. By Claim~\ref{claim:degree_at_least_4}, we have $d(v) \geq 4$ so $N(u) \cap V_3 = \{t_1,t_2\}$.
        Let $\phi$ and $\psi$ be two $2$-dicolourings given by Claim~\ref{claim:existence_of_phi_and_psi}.
        
        Since $t_1$ and $t_2$ are $3$-vertices, Claim~\ref{claim:degree_at_least_4} yields $d^-(t_1) = d^-(t_2) = 1$ because $u$ cannot be their lonely neighbour.
        By Claim~\ref{claim:existence_of_phi_and_psi}, neither $u$ nor $v$ can be recoloured in $\phi$. Thus, for each $j\in [2]$, there is a directed $(v,u)$-path $P_j$ whose internal vertices are all coloured $j$. In particular, we have $\phi(t_1) \neq \phi(t_2)$ because one of $\{t_1,t_2\}$ belongs to $P_1$ and the other one belongs to $P_2$.
        Without loss of generality, we may assume that both $u$ and $t_1$ are coloured 1, and both $v$ and $t_2$ are coloured 2.

\medskip
        
        $(i)$. Let us first show that $t_1$ is blocked in $\phi$. 
        Towards this, consider the sequence successively recolouring $t_1$, $v$, $u$ and $t_1$ again.
        For this not to be a redicolouring sequence from $\phi$ to $\psi$, which would contradict our assumption, it must create a monochromatic directed cycle at one of the steps.
        Assume that $t_1$ is recolourable, that is, the first step of the sequence does not create a monochromatic directed cycle.
        Now, since both $t_1,t_2$ are coloured $2$ at this point, $v$ can also be recoloured. Indeed, any monochromatic directed cycle resulting from setting $v$ to colour $1$ would already be a monochromatic directed cycle in $\psi$. Then, $u$ is coloured the same as its lonely neighbour $v$, allowing us to set the colour of $u$ to $2$, then to recolour $t_1$ and obtain $\psi$.
        Therefore, $t_1$ is blocked in $\phi$. Then, we must have $P_1 = (v,t_1,u)$, otherwise the only in-neighbour of $t_1$ would be coloured 1, and there is a directed $(t_1,v)$-path $P_3$ with internal vertices coloured 2.
        In particular, the out-neighbour of $t_1$ distinct from $u$ is coloured $2$.

        We now consider $t_2$.
        The existence of $P_2$ ensures that $t_2$ has its unique in-neighbour coloured 2, therefore $t_2$ is not blocked to colour 2 in $\phi$.
        Thus we recolour $t_2$ with colour 1, which allows us to then recolour $u$ with colour 2 since its in-neighbourhood is coloured 1. In the resulting dicolouring $\xi$, if $v$ is recolourable, then we can successively recolour $v$ to colour 1 and $t_2$ to colour 2. This gives a redicolouring sequence between $\phi$ and $\psi$, a contradiction. Thus, recolouring $v$ in $\xi$ must create a monochromatic directed cycle. Since $\xi$ agrees with $\psi$ on $V(G) \setminus \{v,t_2\}$, this implies that $P_2=(v,t_2,u)$ and there is a directed $(t_2, v)$-path $P_4$ in $\vec{G}-u$ with internal vertices coloured 1 (in both $\psi$ and $\psi$).

        The existence of $P_1$, $P_2$, $P_3$, and $P_4$ ensures that $d(v) \geq 5$ (see Figure~\ref{fig:not-mixing-1}), and completes the proof of  $(i)$.
        
        \begin{figure}[H]
            \begin{minipage}{\linewidth}
                \begin{center}	
                  \begin{tikzpicture}[thick,scale=1, every node/.style={transform shape}]
            	    \tikzset{vertex/.style = {circle,fill=black,minimum size=5pt,
                                        inner sep=0pt}}
                    \tikzset{edge/.style = {->,> = latex'}}
            	    
                    \node[vertex,red, label=left:$u$] (u) at  (-1,0) {};
                    \node[vertex,red, label=left:$t_1$] (t1) at  (-1,1) {};
                    
                    \node[blue] (q) at  (0,1) {$P_3$};
                    \node[vertex,blue, label=right:$y_1$] (y1) at  (1,1) {};
                    
                    \node[vertex,blue, label=left:$t_2$] (t2) at  (-1,-1) {};
                    \node[vertex,blue, label=right:$v$] (v) at  (1,0) {};
                    
                    \node[red] (q2) at  (0,-1) {$P_4$};
                    \node[vertex,red, label=right:$y_2$] (y2) at  (1,-1) {};
                    
                    \draw[edge] (u) to (v);
                    \draw[edge] (v) to (t1);
                    \draw[edge] (v) to (t2);
                    
                    \draw[edge] (y1) to (v);
                    \draw[edge] (q) to (y1);
                    \draw[edge] (t1) to (q);
                    
                    \draw[edge] (t1) to (u);
                    \draw[edge] (t2) to (u);
                    
                    \draw[edge] (t2) to (q2);
                    \draw[edge] (q2) to (y2);
                    \draw[edge] (y2) to (v);
                  \end{tikzpicture}
              \caption{A $3$-vertex $u$ and its lonely neighbour $v$ in a dicolouring $\phi$ from which we cannot reach the dicolouring $\psi$ swapping the colours of $u$ and $v$ via a redicolouring sequence.
              Vertices coloured $1$ are shown in red, while vertices coloured $2$ are shown in blue.
              Vertex $t_1$ must be blocked, yielding the existence of monochromatic path $P_3$ and the arc $vt_1$. Vertex $v$ must be blocked after recolouring $t_2$, yielding $P_4$ and the arc $vt_2$.} 
              \label{fig:not-mixing-1}
            \end{center}    
          \end{minipage}
        \end{figure}
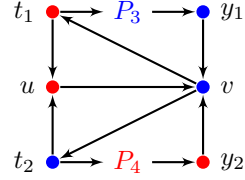
        
        \medskip

        $(ii)$
        Assume $d(v) = 5$.
        Let $y_1$ (resp. $y_2$) be the in-neighbour of $v$ in $P_3$ (resp. $P_4$).
        Recall also that, without loss of generality, we can take $\phi(u)=1$ and $\phi(v)=2$.
        We shall prove that $y_1$ and $y_2$ both have degree at least $4$ in $\vec{G}$.
        \begin{itemize}
            \item Assume for a contradiction that $d(y_1)=3$. 
            If the in-neighbour of $y_1$ on $P_3$ is coloured $2$, we can recolour $y_1$ to $1$ without creating a monochromatic directed cycle coloured $1$. If not, the in-neighbour must be $t_1$, and since its lonely neighbour $v$ is coloured $2$, we can again recolour $y_1$ without creating a monochromatic directed cycle.
            Having recoloured $y_1$, the whole in-neighbourhood of $v$ is now coloured 1. Since $v$ is the lonely neighbour of $t_1$, we can set $t_1$ to colour 2 without creating any monochromatic directed cycle.
            Next we can successively recolour $v$ to $1$ since its out-neighbours are coloured $2$, then $u$ to $2$ since its lonely out-neighbour $v$ is now coloured $1$.
            Now if recolouring $y_1$ yields a monochromatic directed cycle $C$, it does not contain neither $u$ nor $t_1$ because they are both coloured 2 and $v$ is coloured 1. Then $C$ must have already been a monochromatic directed cycle in $\phi$, which is a contradiction.
            Hence we can recolour $y_1$, and finally recolour $t_1$ to obtain a redicolouring sequence from $\phi$ to $\psi$, a contradiction. 
            
            \item Assume for a contradiction that $d(y_2)=3$.
            We start by recolouring $t_2$ to colour 1, and $u$ to colour 2.
            Now we can recolour $y_2$ to colour 2 since $d(y_2)= 3$, the in-neighbour of $y_2$ in $P_4$ is coloured 1 and the out-neighbourhood of $v$ is coloured 1.
            Then, we can recolour $v$ to colour 1 because its in-neighbourhood is coloured 2. Next we can recolour $t_2$ to colour 2 since its in-neighbour $v$ is coloured 1, and we can finally recolour $y_2$ to obtain a redicolouring sequence from $\phi$ to $\psi$, a contradiction.
        \end{itemize}
 \vspace*{-12pt}   \end{proofclaim}
    
    From now on, for each $(\geq 4)$-vertex $v$, we let $S_v$ be the set of $3$-vertices $x$ such that $v$ is the lonely neighbour of $x$.
    
    \begin{claim}\label{claim:boundSv}
    Let $v$ be a $(\geq 4)$-vertex, then  $|S_v| \leq d(v) - 2$.
    \end{claim}
    
    \begin{proofclaim}
        Assume for a contradiction that $|S_v| \geq d(v) - 1$. 
        If $|S_v| = d(v)-1$, we let $w$ be the only neighbour of $v$ which does not belong to $S_v$.
        By directional duality, we may assume that $w$ is an out-neighbour of $v$.
        If $|S_v| = d(v)$, let $w$ be any out-neighbour of $v$. 
        
        We shall find a redicolouring sequence from $\alpha$ to $\beta$.
        Without loss of generality, we may assume $\beta(v)=1$. 
        We first find a redicolouring sequence from $\alpha$ to a $2$-dicolouring $\tilde{\alpha}$ such that $\tilde{\alpha}(v) =1$.
        If $\alpha(v)=1$, there is nothing to do.
        Assume now that $\alpha(v) =2$.  We distinguish two cases, depending on the colour of $w$.
        \begin{itemize}
            \item If $\alpha(w) = 1 \neq \alpha(v)$, then we can set every vertex in $N^+(v)$ to colour 1 without creating any monochromatic directed cycle. Next we can set every vertex in $N^-(v)$ to colour 2. Finally we can recolour $v$ to $1$.
            \item If $\alpha(w) = 2 = \alpha(v)$, then we set every vertex in $N^-(v)$ to colour 1. Next we set every vertex in $N^+(v)$ to colour 2. Finally we can recolour $v$ to $1$.
        \end{itemize}
        
        We now have a $2$-dicolouring $\tilde{\alpha}$ such that $\tilde{\alpha}(v) = \beta(v) = 1$.
        First we set each vertex in $S_v$ to colour 2. Then we consider $\vec{G}' = \vec{G} - (\{v\} \cup S_v)$.
        By the minimality of $\vec{G}$, there is a redicolouring sequence from $\tilde{\alpha}'$ to $\beta'$, the the restrictions of $\tilde{\alpha}$ and $\beta$ to $\vec{G}'$. This sequence extends directly to $\vec{G}$, since $v$ is coloured 1 and every vertex in $S_v$ is coloured 2. Finally we only have to set each vertex $x\in S_v$ to colour $\beta (x)$. This operation does not create any monochromatic directed cycle, because such a cycle $C$ should be coloured 1 (since we only recolour some vertices of $S_v$ to colour 1). But then, every vertex in $C$ is also coloured 1 in $\beta$, a contradiction.
        
        We finally get a redicolouring sequence from $\tilde{\alpha}$ to $\beta$. Since we described above a redicolouring sequence from $\alpha$ to $\tilde{\alpha}$, there is a redicolouring sequence from $\alpha$ to $\beta$, a contradiction.
    \end{proofclaim}
    
    \begin{claim}\label{claim:degree4-Sv0}
        Let $v$ be a 4-vertex. If $|S_v| = 0$, then $|N(v) \cap V_3| \leq 2$.
    \end{claim}

    \begin{proofclaim}
        Assume for a contradiction that $|S_v| = 0$ and  $|N(v) \cap V_3| \geq 3$. 
        We consider $\vec{G}' = \vec{G} - (\{v\} \cup (N(v) \cap V_3))$.
        By minimality of $\vec{G}$, there is a redicolouring sequence $\gamma'_0,\dots,\gamma'_r$ where $\gamma'_0$ and $\gamma'_r$ are the restrictions of $\alpha$ and $\beta$ to $\vec{G}'$. 
        In order to get a contradiction, we exhibit a redicolouring sequence from $\alpha = \gamma_0$ to $\beta = \gamma_r$ in $\vec{G}$ as follows.
        For $i=1$ to $r$, we show that there exists a $2$-dicolouring $\gamma_i$ of $\vec{G}$ such that $\gamma'_i$ is the restriction of $\gamma_i$ on $\vec{G}'$, and there is a redicolouring sequence from $\gamma_{i-1}$ to $\gamma_i$. The concatenation of these sequences will then yield the desired sequence between $\alpha$ and $\beta$.

        Assume that for any $i \geq 1$, we have defined dicolouring $\gamma_{i-1}$ such that $\gamma'_{i-1}$ is the restriction of $\gamma_{i-1}$ to $\vec{G}'$, let us and exhibit a redicolouring sequence from $\gamma_{i-1}$ to some $\gamma_i$ such that $\gamma'_i$ is a restriction of $\gamma_i$ to $\vec{G}'$.
        In $\gamma_{i-1}$, we first recolour each vertex of $N(v) \cap V_3$ with the colour different from the one of its lonely neighbour to get a new $2$-dicolouring $\tilde{\gamma}_{i-1}$ (possibly equal to $\gamma_{i-1}$).
        Note that $\gamma'_{i-1}$ is still the restriction of $\tilde{\gamma}_{i-1}$ on $\vec{G}'$. 
        Let $x_i$ be the vertex recoloured between $\gamma'_{i-1}$ and $\gamma'_i$.
        
        If recolouring $x_i$ in our current dicolouring $\tilde{\gamma}_{i-1}$ of $\vec{G}$ creates no monochromatic directed cycle, we let $\gamma_i$ be the resulting dicolouring.
        Then, we have defined a sequence from $\gamma_{i-1}$ to $\tilde{\gamma}_{i-1}$ to $\gamma_i$, yielding the desired property.
        If the recolouring of $x_i$ in $\tilde{\gamma}_{i-1}$ does yield a monochromatic directed cycle, the cycle must contain at least one vertex in  $\{v\} \cup (N(v) \cap V_3$).
        Moreover, it must always contain a 3-vertex $x$ that is a neighbour of $v$, because when it contains $v$ it must also contain two of its neighbours, and by assumption at most one is not a $3$-vertex.
        Since in $\tilde{\gamma}_{i-1}$, $x$ is coloured differently from its lonely neighbour, $x_i$ must be the lonely neighbour of $x$. 
        We may assume without loss of generality that $x$ is an out-neighbour of $v$.
        Since all $3$-vertices in $N(v)$, that is, all but at most one vertex in $N(v)$, are now coloured differently from their lonely neighbour, we can recolour $v$ without creating any monochromatic directed cycle.
        Hence, up to this recolouring, we may assume that the two in-neighbours of $x$ have the same colour.
        If this colour is the same as $x$, then we can recolour $x$ and then recolour $x_i$, otherwise we can directly recolour $x_i$. In both cases, we let $\gamma_i$ be the resulting dicolouring, which has been obtained by a redicolouring sequence from $\gamma_{i-1}$.
        
        Concatenating the sequences $\gamma_{i-1},...,\gamma_{i}$ for $i \in [r]$ yields a sequence from $\alpha$ to a dicolouring $\gamma_r$ which agrees with $\beta$ on $V(\vec{G}')$.
        To extend this into a redicolouring sequence to $\beta$, in $\gamma_r$ we start by recolouring each vertex of $N(v) \cap V_3$ with the colour different from the one of its lonely neighbour.
        Then, if necessary, we can recolour $v$ to $\beta(v)$ without creating any monochromatic directed cycle. Finally we recolour in any order the neighbours of $v$ that need to be recoloured with their colour in $\beta$.
        If recolouring one of these vertices, say $y$, were to create a monochromatic directed cycle $C$, since $\beta$ is a 2-dicolouring, $C$ would have to contain both $y$ and a neighbour of $v$ coloured differently than in $\beta$. But these neighbours can only be other $3$-vertices that we have not recoloured at this point, which are therefore coloured differently from their lonely neighbour, thus cannot be part of a monochromatic directed cycle.
        Hence, we have a redicolouring sequence from $\alpha$ to $\beta$, yielding the contradiction.
    \end{proofclaim}

    \begin{claim}\label{claim:degree4-Sv1}
        Let $v$ be a 4-vertex. If $|S_v| = 1$, then $|N(v) \cap V_3| \leq 2$.
    \end{claim}
    \begin{proofclaim}
        Assume for a contradiction that $|S_v| = 1$ and $|N(v) \cap V_3| \geq 3$, and let $u$ be the only vertex in $S_v$.
        By directional duality, we may assume that $u$ is an in-neighbour of $v$.
        Note that by Claim~\ref{claim:degree_at_least_4}, $d^-(v) = d^+(v) = 2$.
        This implies that $v$ has at least one out-neighbour $x$ in $V_3$, and $d^+(x) = 1$ since $S_v = \{u\}$.
        
        By Claim~\ref{claim:existence_of_phi_and_psi}, there are two $2$-dicolourings  $\phi$ and $\psi$ of $\vec{G}$ that agree on $V(\vec{G}) \setminus \{u,v\}$ and such that there is no redicolouring sequence from $\phi$ to $\psi$.
        In this case, $\phi(u) \neq \phi(v)$ and both $u$ and $v$ are blocked in $\phi$, for otherwise we could recolour them one after the other to yield $\psi$.
        Without loss of generality, let $\phi(u) = 1$ and $\phi(v) = 2$. 
        For each $j\in [2]$ there is a $(v,u)$-path $P_j$ whose internal vertices are all coloured $j$ in $\phi$. We now distinguish two cases, depending on whether $x$ belongs to $P_1$ or $P_2$.
        \begin{itemize}
            \item Assume first that $x$ is the out-neighbour of $v$ in $P_1$.
            Then $x$ can be recoloured since its only out-neighbour is coloured 1.
            After recolouring $x$, every out-neighbour of $v$ is coloured 2, hence we can successively recolour $v$ to 1 and $u$ to 2. We finally can recolour $x$ to 1, since this directly gives dicolouring $\psi$, yielding a redicolouring sequence from $\phi$ to $\psi$, a contradiction.
            
            \item Assume now that $x$ is the out-neighbour of $v$ in $P_2$.
            Then, $x$ can be recoloured since its only out-neighbour is either coloured $2$ or it is $u$ (and the only out-neighbour of $u$ is coloured $2$).
            After recolouring $x$, every out-neighbour of $v$ has colour 1, and we can recolour $u$ to colour 2. Now, recolouring $v$ to colour 1 does not create any monochromatic directed cycle.
            Indeed, such a cycle $C$ would necessarily contain $x$, otherwise $C$ would already be monochromatic in $\psi$, and be coloured 1.
            Note that $x$ has only one out-neighbour, which is necessarily coloured $2$, because it is either internal to $P_2$, or it is $u$, which is now coloured $2$. 
            We finally can recolour $x$, yielding a redicolouring sequence from $\phi$ to $\psi$, a contradiction.
        \end{itemize}
    \end{proofclaim}

    \begin{claim}\label{claim:degree4-Sv2}
        Let $v$ be a 4-vertex with $|S_v| = 2$. Set $S_v = \{u_1,u_2\}$ and $N(v) \setminus S_v = \{w_1,w_2\}$. Then each of the following holds:
        \begin{itemize}
            \item[(i)] $v$ has exactly one in-neighbour and one out-neighbour in $S_v$;
            \item[(ii)] $v$ has exactly one in-neighbour and one out-neighbour in $\{w_1,w_2\}$;
            \item[(iii)] $\vec{G}$ has a 2-dicolouring $\gamma$ such that $\gamma(u_1) = \gamma(u_2) = 1$, $\gamma(v) = \gamma(w_1) = \gamma(w_2) = 2$ and there exists no redicolouring sequence from $\gamma$ that recolours at least one vertex in $N[v]$;
            \item[(iv)] $d(w_1) \geq 4$ and $d(w_2) \geq 4$.
        \end{itemize}
    \end{claim}

    \begin{proofclaim}
        Figure~\ref{fig:claim_169} illustrates the neighbourhood of such a $4$-vertex $v$ coloured with $\gamma$.

        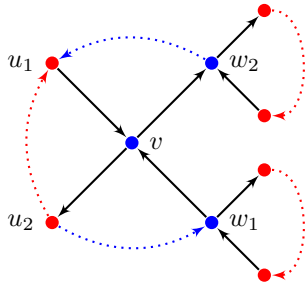
\begin{figure}[H]
            \begin{minipage}{\linewidth}
                \begin{center}	
                  \begin{tikzpicture}[thick,scale=1, every node/.style={transform shape}]
            	    \tikzset{vertex/.style = {circle,fill=black,minimum size=5pt,
                                        inner sep=0pt}}
                    \tikzset{edge/.style = {->,> = latex'}}
            	    
                    \node[vertex,blue, label=right:$v$] (v) at  (0,0) {};
                    \node[vertex,blue, label=right:$w_1$] (w1) at  (-45:1.5) {};
                    \node[vertex,blue, label=right:$w_2$] (w2) at  (45:1.5) {};
                    \node[vertex,red, label=left:$u_1$] (u1) at  (135:1.5) {};
                    \node[vertex,red, label=left:$u_2$] (u2) at  (-135:1.5) {};
                    \draw[edge] (v) to (w2) {};
                    \draw[edge] (v) to (u2) {};
                    \draw[edge] (w1) to (v) {};
                    \draw[edge] (u1) to (v) {};

                    \node[vertex, red, xshift=0.7cm, yshift=0.7cm] (nbr1w1) at  (-45:1.5) {};
                    \node[vertex, red, xshift=0.7cm, yshift=-0.7cm] (nbr2w1) at  (-45:1.5) {};
                    \draw[edge] (w1) to (nbr1w1) {};
                    \draw[edge] (nbr2w1) to (w1) {};

                    \node[vertex, red, xshift=0.7cm, yshift=0.7cm] (nbr1w2) at  (45:1.5) {};
                    \node[vertex, red, xshift=0.7cm, yshift=-0.7cm] (nbr2w2) at  (45:1.5) {};
                    \draw[edge] (w2) to (nbr1w2) {};
                    \draw[edge] (nbr2w2) to (w2) {};
                    
                    \draw[edge,dotted,red] (u2) to[out=120, in=-120] (u1) {};
                    \draw[edge,dotted,blue] (u2) to[out=-30, in=-150] (w1) {};
                    \draw[edge,dotted,blue] (w2) to[out=150, in=30] (u1) {};
                    
                    \draw[edge,dotted,red] (nbr1w1) to[out=0, in=0] (nbr2w1) {};
                    
                    \draw[edge,dotted,red] (nbr1w2) to[out=0, in=0] (nbr2w2) {};
                    
                  \end{tikzpicture}
              \caption{The neighbourhood of a $4$-vertex $v$ with $|S_v| = 2$ coloured with the particular dicolouring $\gamma$ defined in Claim~\ref{claim:degree4-Sv2}. Vertices coloured $1$ are shown in red, while vertices coloured $2$ are shown in blue. The existence of the coloured dotted paths is guaranteed because no vertex of $N[v]$ is recolourable in $\gamma$.} 
              \label{fig:claim_169}
            \end{center}    
          \end{minipage}
        \end{figure}
    
        $(i)$ By Claim~\ref{claim:degree_at_least_4}, we have $d^+(v) = d^-(v) = 2$.
        Suppose $(i)$ does not hold, and assume without loss of generality that both $u_1$ and $u_2$ are in-neighbours of $v$, so both $w_1$ and $w_2$ are out-neighbours of $v$.
        Towards a contradiction, we exhibit a redicolouring sequence from $\alpha$ to $\beta$. Without loss of generality, we assume $\beta(v)=1$.
        
        Starting from $\alpha$, we first show how to obtain a dicolouring where $v$ is coloured $1$. 
        If $\alpha(v) = 1$, there is nothing to do.
        If $\alpha(v) = 2$, we distinguish three cases depending on the colours of $\alpha(w_1)$ and $\alpha(w_2)$:
        \begin{itemize}
            \item If $\alpha(w_1) = \alpha(w_2) = 2 = \alpha(v)$, then we can directly recolour $v$.
            \item If $\alpha(w_1) = \alpha(w_2) = 1 \neq \alpha(v)$, then we can set both $u_1$ and $u_2$ to $2$, and then recolour $v$ with $1$.
            \item Else, $\alpha(w_1) \neq \alpha(w_2)$.
             Since $\alpha(v) = 2$, we first set $u_1$ and $u_2$ to colour $1$, naming the resulting $2$-dicolouring $\hat{\alpha}$.
             Then, we consider $\vec{G}' = \vec{G} - \{v,u_1,u_2\}$ and let $\hat{\alpha}'$ be the restriction of $\hat{\alpha}$ to $\vec{G}'$.
             By minimality of $\vec{G}$, $\vec{G}'$ is 2-mixing.
             In particular, there is a redicolouring sequence from $\hat{\alpha}'$ to its mirror.
             Along this sequence, since we initially set $\hat{\alpha}'(w_1) \neq \hat{\alpha}'(w_2)$, there is a 2-dicolouring $\breve{\alpha}'$ of $\vec{G}'$ such that $\breve{\alpha}'(w_1) = \breve{\alpha}'(w_2)$.
             The redicolouring sequence from $\hat{\alpha}'$ to $\breve{\alpha}'$ in $\vec{G}'$ directly extends to a redicolouring sequence from $\hat\alpha$ to a $2$-dicolouring $\breve{\alpha}$ in $\vec{G}$.
             Indeed, in $\hat\alpha$, both $u_1$ and $u_2$ are coloured differently from their lonely neighbour $v$, and $v$ has no other in-neighbours, thus applying the same sequence in $\hat\alpha$ does not create any monochromatic directed cycle.
             After this recolouring, we obtain dicolouring $\breve\alpha$ such that $\breve{\alpha}(w_1) = \breve{\alpha}(w_2)$, allowing us to apply one of the first two items to recolour $v$ with 1.
        \end{itemize}
         We have shown that there is a (possibly empty) redicolouring sequence from $\alpha$ to a $2$-dicolouring $\alpha^*$ such that $\alpha^*(v) = \beta(v) = 1$.
         We now start from $\alpha^*$, set both $u_1$ and $u_2$ to $2$, and name the resulting $2$-dicolouring $\tilde{\alpha}$.
         Consider $\vec{G}' = \vec{G} - \{v,u_1,u_2\}$ and $\tilde{\alpha}'$, $\beta'$ the restrictions of $\tilde{\alpha}$ and $\beta$ to $\vec{G}'$.
         By minimality of $\vec{G}$, there is a redicolouring sequence from $\tilde{\alpha}'$ to $\beta'$ in $\vec{G}'$.
         This redicolouring sequence immediately extends to $\vec{G}$, by keeping $\tilde{\alpha}$ on $v,u_1,u_2$, and we may only need to change the colour of $u_1$ and $u_2$ at the end.
         Therefore there is a redicolouring sequence from $\alpha$ to $\beta$, a contradiction.
         This proves $(i)$.

\medskip         
          
    $(ii)$ follows directly from $(i)$ by Claim~\ref{claim:degree_at_least_4}.

\medskip

        $(iii)$ By $(i)$ and $(ii)$, we may assume that $u_1$ and $w_1$ are the in-neighbours of $v$ and $u_2$ and $w_2$ are the out-neighbours of $v$.
        Assume for the sake of contradiction that $(iii)$ does not hold, that is:
        \begin{center}
             ($\star$) {\it for any 2-dicolouring $\xi$ of $\vec{G}$, such that $\xi(u_1) = \xi(u_2)$ and $\xi(v) = \xi(w_1) = \xi(w_2)$ with $\xi(v) \neq \xi(u_1)$, there is a redicolouring sequence from $\xi$ to a 2-dicolouring $\xi'$ such that for at least one vertex $x$ in $N[v]$, $\xi(x) \neq \xi'(x)$.} 
        \end{center}
        Assuming this, we will prove that there is a redicolouring sequence from $\alpha$ to $\beta$, which is a contradiction, showing the existence of $\gamma$. 
        
        First we show that there is a redicolouring sequence from $\alpha$ to some dicolouring $\Tilde{\alpha}$, where $\Tilde{\alpha}(w_1) \neq \Tilde{\alpha}(w_2)$. We assume without loss of generality that $\alpha(v) = 1$.
        If $\alpha(w_1) \neq \alpha(w_2)$, we let $\Tilde\alpha=\alpha$, otherwise, $\alpha(w_1)=\alpha(w_2)$ and we distinguish two cases according to the colour of $\alpha(w_1)$.
        \begin{itemize}
            \item Assume $\alpha(w_1) = \alpha(w_2)  = 2 \neq \alpha(v)$. 
            Starting from $\alpha$, we first set both $u_1$ and $u_2$ to $2$.
            Then, we consider $\vec{G}' = \vec{G} - \{u_1,u_2,v\}$, which must be $2$-mixing by minimality of $\vec{G}$.
            In particular, there is a redicolouring sequence in $\vec{G}'$ from $\alpha'$, the restriction of $\alpha$ to $\vec{G}'$, to its mirror $\overline{\alpha'}$.
            Along this sequence, we let $\Tilde{\alpha}'$ be the first dicolouring where $w_1$ and $w_2$ are coloured differently.
            The redicolouring sequence from $\alpha'$ to $\Tilde{\alpha}'$ in $\vec{G}'$ directly extends to a redicolouring sequence from $\alpha$ to $\Tilde{\alpha}$ in $\vec{G}$.
            Indeed, we keep $v$ coloured $1$ and $u_1$ and $u_2$ coloured $2$ throughout the sequence, which does not create any monochromatic directed cycle since $v$ does not have any neighbour coloured $1$ until reaching $\Tilde{\alpha}$, when it has exactly one.
            
            \item Assume $\alpha(w_1) = \alpha(w_2) = 1 = \alpha(v)$.
            Starting from $\alpha$, we first set both $u_1$ and $u_2$ to $2$.
            Now by ($\star$) there is a redicolouring sequence from $\alpha$ to a 2-dicolouring $\hat{\alpha} $ where one vertex $x$ in $N[v]$ has a different colour.  
            If $x$ is $w_1$ or $w_2$, then we are done.
            If $x$ is $v$, then we are done by the previous case. Finally if $x$ is $u_1$ or $u_2$, then $v$ has three neighbours coloured $1$, hence $v$ can be recoloured $2$, and we are done by the previous case (swapping the roles of colours 1 and 2).
        \end{itemize}
        The same proof applies to $\beta$ to yield a redicolouring sequence from $\beta$ to some $2$-dicolouring $\Tilde{\beta}$ such that $\Tilde{\beta}(w_1) \neq \Tilde{\beta}(w_2)$.
        It now suffices to exhibit a redicolouring sequence from $\Tilde\alpha$ to $\Tilde\beta$.
        
        Consider $\vec{G}' = \vec{G} - \{v,u_1,u_2\}$, which is $2$-mixing by minimality of $\vec{G}$, giving us a redicolouring sequence $\tilde{\alpha}'=\eta_0',\dots, \eta_r'=\tilde{\beta}'$, where $\tilde{\alpha}'$ and $\tilde{\beta}'$ are the restrictions of $\tilde{\alpha}$ and $\tilde{\beta}$ to $\vec{G}'$.
        We will extend this redicolouring sequence to $\vec{G}$.
        To do so, for $i=1$ to $r$, we show that there exists a $2$-dicolouring $\eta_i$ of $\vec{G}$ such that $\eta'_i$ is the restriction of $\eta_i$ on $\vec{G}'$, and there is a redicolouring sequence from $\eta_{i-1}$ to $\eta_i$.
        To ensure that extending this sequence does not create any monochromatic directed cycle, we will maintain the property that $\{v,w_1,w_2\}$ is not monochromatic in each $\eta_i$.
        
        We first set $\eta_0$ to $\tilde{\alpha}$.
        Since $\tilde{\alpha}(w_1) \neq \tilde{\alpha}(w_2)$, $\{v,w_1,w_2\}$ is not monochromatic in $\eta_0$.
        Assume now that $\eta_{i-1}$ exists, and let us define $\eta_i$.
        Let $x_i$ be the vertex that is recoloured from $\eta_{i-1}'$ to $\eta_{i}'$.
        Without loss of generality, say $v$ is coloured $1$ in $\eta_{i-1}$.
        We may assume that $\eta_{i-1}(u_1) = \eta_{i-1}(u_2) = 2$, otherwise we may recolour them.
        Then, if recolouring $x_i$ does not break the invariant that $\{v,w_1,w_2\}$ is not monochromatic, we can recolour $x_i$ without creating any monochromatic directed cycle and obtain the desired $\eta_i$.
        Indeed, a resulting monochromatic directed cycle cannot contain $u_1$ or $u_2$, so it must contain $v$ and therefore both $w_1$ and $w_2$, which are coloured differently.  
        If recolouring $x_i$ breaks the invariant, $x_i$ must belong to $\{w_1,w_2\}$.
        Assume then that $x_i = w_1$, and recolouring $w_1$ breaks the invariant.
        This implies that $w_2$ is currently coloured 1 and $w_1$ is going to be recoloured from colour 2 to colour 1.
        In this case, all in-neighbours of $v$ are coloured $2$, allowing us to first recolour $u_2$ without creating any monochromatic directed cycle.
        Now, all out-neighbours of $v$ are coloured $1$, letting us set $v$ to colour 2.
        We finally can recolour $w_1$ to colour 1 and we get the desired $\eta_{i}$ extending $\eta'_{i}$ and preserving the invariant. The arguments hold symmetrically when the recoloured vertex $x_i$ is $w_2$.
        
        Concatenating the resulting sequences $\eta_{i-1},...,\eta_i$ for $i \in [1,r]$, we obtain a redicolouring sequence from $\tilde{\alpha}$ to a $2$-dicolouring $\beta^*$ which agrees with $\Tilde{\beta}$ on $V(\vec{G}')$. Moreover, $\{v,w_1,w_2\}$ is not monochromatic in $\beta^*$.
        We now turn to exhibiting a redicolouring sequence from $\beta^*$ to $\tilde{\beta}$.
        Let us assume, without loss of generality, that $\beta^*(v)=1$. We may first set $u_1$ and $u_2$ to colour $2$, as $v$ is their lonely neighbour.
        We show how to recolour $v$ to colour $2$, when required, without affecting the colours of $V(\vec{G}')$. Note that $\tilde{\beta}$ is a dicolouring, so any monochromatic directed cycle resulting from recolouring $v$ must use one of $u_1$ or $u_2$.
        We distinguish two cases depending on the colour of $\Tilde{\beta}(w_1)$.
         \begin{itemize}
             \item If $\Tilde{\beta}(w_1)=1$, we know by assumption on $\Tilde{\beta}$ that $\Tilde{\beta}(w_2)=2$.
             Now, all out-neighbours of $v$ are coloured $2$, letting us first recolour $u_1$ to $1$.
             Then, all in-neighbours of $v$ are coloured $1$, letting us set $v$ to colour $2$.
             \item If $\Tilde{\beta}(w_1)=2$, we know by assumption that $\Tilde{\beta}(w_2)=1$.
             Symmetrically, all in-neighbours of $v$ are coloured $2$, letting us recolour $u_2$ to $1$.
             Then, all out-neighbours of $v$ are coloured $1$, letting us recolour $v$ to $2$.
         \end{itemize}
        Finally, if necessary, we may recolour $u_1$ and $u_2$. To do so, we may initially set $u_1$ and $u_2$ to the colour different from $\Tilde{\beta}(v)$. Then, if only one of $u_1,u_2$ needs to be recoloured, this can be done since $\Tilde\beta$ is a dicolouring. If both need to be recoloured, this can also be done in any order, as any resulting monochromatic directed cycle would also exist in $\Tilde\beta$.
        This yields redicolouring sequence from $\tilde{\alpha}$ to $\tilde{\beta}$, which together with the redicolouring sequences from $\alpha$ to $\tilde{\alpha}$ and $\beta$ to $\tilde{\beta}$ yields a redicolouring sequence from $\alpha$ to $\beta$, a contradiction.
        This achieves to prove $(iii)$.
        
  \medskip
        
        $(iv)$ This is a consequence of $(iii)$, let us consider the dicolouring $\gamma$ given by $(iii)$ with $\gamma(u_1)=\gamma(u_2)=1$ and $\gamma(v)=\gamma(w_1)=\gamma(w_2) = 2$.
        In $\gamma$, $u_2$ cannot be recoloured to $2$, therefore there is a $(u_2, w_1)$-path with internal vertices coloured $2$.
        If $d(w_1) = 3$, then $w_1$ could be recoloured $1$ because its out-neighbour $v$ is coloured $2$ and one of its in-neighbours is either coloured $2$ or is $u_2$, whose unique in-neighbour $v$ is coloured $2$.
        This would contradict $(iii)$, hence $d(w_1)\geq 4$.
        We can show symmetrically that $d(w_2)\geq 4$, achieving to prove $(iv)$.
    \end{proofclaim}
    
    \begin{claim}\label{claim:degree4_V3_atmost2}
        If $v$ is a 4-vertex, then $|N(v) \cap V_3| \leq 2$.
    \end{claim}
    
    \begin{proofclaim}
        This is a direct consequence of Claims~\ref{claim:boundSv},~\ref{claim:degree4-Sv0},~\ref{claim:degree4-Sv1} and~\ref{claim:degree4-Sv2}.
    \end{proofclaim}
    
    We shall now use the Discharging Method. The initial charge of each vertex $x$ is $d(x)$, and we apply the following rules: 
    
    \begin{itemize}
        \item[(R1)] Each vertex $x$ such that $d(x)=3$ and $|N(x) \cap V_3| \geq 2$ receives $\frac{1}{2}$ charge from its lonely neighbour.

        \item[(R2)] Each vertex $x$ such that $d(x)=3$ and $|N(x) \cap V_3| \leq 1$ receives $\frac{1}{4}$ from each of its $(\geq 4)$-neighbours.
    \end{itemize}
    
    Let us now show that the final charge $w^*(x)$ of a vertex $x$ in $\vec{G}$ is at least $\frac{7}{2}$, showing that $|A| \geq \frac{7}{4}|V| $, which contradicts the assumption $|A| < \frac{7}{4}|V|$.
    
    \begin{itemize}
        \item Assume $d(x) = 3$ and $|N(u) \cap V_3| \geq 2$, then $x$ receives $\frac{1}{2}$ by (R1). Hence $w^*(x) = d(x) + \frac{1}{2} = \frac{7}{2}$. 
        
        \item Assume $d(x) = 3$ and $|N(u) \cap V_3| \leq 1$, then $x$ receives $\frac{1}{4}$ by (R2) at least twice. Hence $w^*(x) \geq d(x) + \frac{1}{2} = \frac{7}{2}$. 
        
        \item Assume $d(x) = 4$. By Claim~\ref{claim:v_has_degree5}, $x$ does not give any charge through (R1), and by Claim~\ref{claim:degree4_V3_atmost2} it gives at most twice $\frac{1}{4}$ by (R2). Hence $w^*(x) \geq d(x) - 2 \cdot \frac{1}{4} = \frac{7}{2}$. 
        
        \item Assume $d(x) = 5$.
        If $x$ gives $\frac{1}{2}$ charge at least once by (R1), then by Claim~\ref{claim:v_has_degree5}, $x$ has at least two neighbours that did not take any charge from $x$. Therefore $w^*(x)\geq  d(x) - 3\times\frac{1}{2} = \frac{7}{2}$. 
        
        If $x$ does not give by (R1), then it gives at most $\frac{1}{4}$ to each of its neighbours. Thus $w^*(x)\geq d(x) - d(x) \times \frac{1}{4} = \frac{15}{4} >  \frac{7}{2}$. 
        
        \item Finally assume $d(x) \geq 6$. By Claim~\ref{claim:boundSv}, $x$ gives $\frac{1}{2}$ by (R1) to at most $d(x) - 2$ of its neighbours. Thus $w^*(x) \geq d(x) - \frac{1}{2}(d(x)-2) - 2\times \frac{1}{4} = \frac{d(x)}{2} + \frac{1}{2} \geq \frac{7}{2}$. 
    \end{itemize}
    This completes the proof of Theorem~\ref{theorem:not-2mixing-density}.
\end{proof}

\section{Further research}
\label{section:open_problems}

In this first paper, we established the first results on digraph redicolouring. This is obviously just the tip of the iceberg and many open questions arise.
Forthwidth, we detail a few of them.

\medskip

In Section~\ref{section:complexity}, we prove 
that {\sc $k$-Dicolouring Path} is PSPACE-complete for all $k\geq 2$.
But we did not prove any complexity result about
{\sc Directed Is $k$-Mixing}.

\begin{problem}
    What is the complexity of {\sc Directed Is $k$-Mixing}?
\end{problem}

We believe that this is PSPACE-hard for all $k\geq 2$.
To settle the complexity of {\sc Directed Is $k$-Mixing}, it could be helpful to settle the complexity of the following particular case of {\sc $2$-Dicolouring Path}. (Recall that the mirror of a $2$-dicolouring $\alpha$ of $D$, is the 2-dicolouring $\bar{\alpha}$ of $D$ such that $\alpha(v) \neq \bar{\alpha}(v)$ for all $v\in V(D)$.)

\smallskip

\defproblem{\sc Mirror-Reachability}{A 2-dicolouring $\alpha$ of a digraph $D$.}{Is there a path between $\alpha$ and its mirror $\bar{\alpha}$ in $\mathcal{D}_2(D)$?}

\begin{problem}
    What is the complexity of {\sc Mirror-Reachability}?
\end{problem}

A particular case of non $k$-mixing digraphs are $k$-freezable digraphs. It would then be interesting to consider the complexity of the following problem.

\defproblem{\sc Directed Is $k$-Freezable}{A $k$-dicolourable digraph $D$.}{Is $D$ $k$-freezable?}

Note that deciding whether a digraph is $k$-freezable is
NP-complete in general since we can reduce easily from $k$-dicolourability for all $k\geq 2$.
Indeed, for a digraph $D$, let $D'$ be the digraph obtained from $D$ by adding on each vertex $v\in V(D)$ a complete bidirected graph $K_k^v$ which contains $v$ and $k-1$ new vertices. Trivially, $D'$ is $k$-dicolourable if and only if $D$ is $k$-dicolourable, and every $k$-dicolouring of $D'$ (if one exists) is necessarily frozen because of the complete bidirected graphs.
Therefore $D'$ is $k$-freezable if and only if $D$ is $k$-dicolourable.

A related problem is the one of deciding whether a vertex is frozen in a given $k$-dicolouring $\alpha$ of a digraph $D$.  Recall that a vertex $v$ is frozen in $\alpha$ if $\beta(v) = \alpha(v)$ for any $k$-dicolouring $\beta$ in the same connected component of $\alpha$ in ${\cal D}_k(D)$.

\medskip

\defproblem{\sc $k$-Frozen Vertex}{A digraph $D$, a $k$-dicolouring $\alpha$ of $D$, and a vertex $v$ of $D$.}{Is $v$ frozen in $\alpha$?}

{\sc $2$-Frozen Vertex} is  PSPACE-complete. This comes from the following result from \cite{hearnTCS343}: given a cubic graph $G$, a mapping $\phi : V(G) \xrightarrow{} \{1,2\}$, a proper orientation $\vec{G}$ of $G$, and an edge $xy$ of $G$, deciding whether there is a reorienting sequence from $\vec{G}$ that reverse $xy$ is PSPACE-complete.
Hence, the same reduction as the one used for Theorem~\ref{thm:pspace-completeness} also yields PSPACE-completeness of {\sc $2$-Frozen Vertex}.

One can then easily derive that {\sc $k$-Frozen Vertex} is  PSPACE-complete for any $k\geq 2$.
Indeed, consider a (non-acyclic) digraph $D_2$ and a  $2$-dicolouring $\alpha_2$ of $D_2$. Let $D_k$ be the digraph obtained the disjoint union of $D_2$ and bidirected complete graph $\bid{K}_{k-2}$ on $k-2$ vertices $y_1, \dots , y_{k-2}$ and adding a digon between any vertex of $D_2$ and any vertex of $\bid{K}_{k-2}$. Let $\alpha_k$ be the $k$-dicolouring of $D_k$ defined by $\alpha_k(v)=\alpha_2(v)$ for all $v\in V(D_2)$ and $\alpha_k(y_i) = i+2$ for every $i\in [k-2]$. One can easily check that a vertex $v$ in $V(D_2)$ is frozen in $\alpha_2$ if and only if it is frozen in $\alpha_k$.

\bigskip

For each of the above problems which are proved or conjectured to be PSPACE-complete, it is natural to determine the smallest possible classes of digraphs on which the problem remains PSPACE-complete. 
For example, in Theorem~\ref{thm:pspace-completeness}, we prove that it remains PSPACE-complete for oriented graphs in $(ii)$ and planar oriented graphs with maximum degree at most 6 in $(iv)$.
This raises the following questions.
\begin{problem}\label{pb:Dic-Path}
\begin{itemize}
    \item[(a)] What is the smallest $D_k$ (resp. $D^*_k$) such that {\sc $k$-Dicolouring Path} remains PSPACE-complete for oriented graphs (resp. planar oriented graphs) with maximum degree at most $D_k$  (resp. $D^*_k$)?
    \item[(b)] Does {\sc $k$-Dicolouring Path} remain 
  PSPACE-complete for digraphs with no directed cycles of length less than $g$, for any fixed $g$.  
    \end{itemize}
\end{problem}

Regarding Problem~\ref{pb:Dic-Path}~(a),  Theorem~\ref{thm:pspace-completeness} and Theorem~\ref{thm:subcubic} imply that $D_2 \in \{4,5\}$. Thus the first question to address is the following.
\begin{problem} 
    What is the complexity of {\sc $2$-Dicolouring Path} restricted to digraphs with maximum degree at most $4$?
\end{problem}

In the opposite direction, it would be interesting to determine the largest possible classes of digraphs for which {\sc $k$-Dicolouring Path} can be solved in polynomial time.
For example, Corollary~\ref{cor:subcubic} implies that {\sc Directed Is $2$-mixing} is polynomial-time solvable for subcubic digraphs.
This raises the following questions.
\begin{problem}\label{pb:Mix-Freez}
\begin{enumerate}[(a)]
    \item What is the largest $M_k$ such that {\sc Directed Is $k$-mixing} is polynomial-time solvable for digraphs with maximum degree at most $M_k$? 
    \item What is the largest $M^{\ast}_k$ such that {\sc Directed Is $k$-freezable} is polynomial-time solvable for digraphs with maximum degree at most $M^{\ast}_k$?
    \end{enumerate}
\end{problem}

A first step towards tackling Problem~\ref{pb:Mix-Freez}~(a) would be to determine whether Theorem~\ref{thm:subcubic} extends to digraphs with maximum degree $k$ for every odd $k$.
\begin{problem}
Let $k=2\ell+1$ be an odd positive integer.
Is every oriented graph with maximum degree $k$ $(\ell+1)$-mixing?
\end{problem}

Corollary~\ref{corollary:kfrozen-density} directly implies that a planar oriented graph is not $3$-freezable. It is then natural to ask whether a planar oriented graph is always $3$-mixing.
\begin{problem}
    Is every planar oriented graph $3$-mixing?
\end{problem}

\medskip

In Sections~\ref{section:diameter} and \ref{section:density}, we established several conditions for a digraph to be $k$-mixing and sometimes some bounds on the diameter.
It is then natural to ask whether the bound in the diameter is tight or not.
For example, Theorem~\ref{thm:min-degen} states that every $(k-2)$-min-degenerate digraph $D$ is $k$-mixing and following its proof one can show that the diameter of ${\cal D}_k(D)$ is at most $2^n$ with $n=|V(D)|$. But it is likely to be smaller, perhaps even polynomial in $n$.

\begin{problem}
What is the maximum diameter of ${\cal D}_k(D)$ over all $(k-2)$-min-degenerate digraphs $D$ of order $n$?
\end{problem}

Another example is the diameter of ${\cal D}_2(\vec{G})$ for a subcubic oriented graph $\vec{G}$. By Theorem~\ref{thm:subcubic}, it is at most $2n$. But, we do not know if this bound is tight.
\begin{problem}
What is the maximum diameter of ${\cal D}_2(\vec{G})$ over all subcubic oriented graphs $\vec{G}$ of order $n$?
\end{problem}

\end{sloppypar}

\bibliography{refs}

\end{document}